\documentclass[12pt,a4paper]{article}
\usepackage[nosort]{cite}
\usepackage[hyperref,bulletsep]{collect}
\usepackage{amsmath,amssymb,graphicx}
\usepackage{pst-all}
\usepackage{bbm}
\usepackage{youngtab}
\usepackage{color}

\let\oldPhi=\Phi
\let\oldPsi=\Psi
\let\oldGamma=\Gamma
\let\oldDelta=\Delta
\let\oldSigma=\Sigma
\let\oldTheta=\Theta
\let\oldPi=\Pi
\let\oldUpsilon=\Upsilon
\renewcommand{\Phi}{\mathnormal{\oldPhi}}
\renewcommand{\Psi}{\mathnormal{\oldPsi}}
\renewcommand{\Gamma}{\mathnormal{\oldGamma}}
\renewcommand{\Sigma}{\mathnormal{\oldSigma}}
\renewcommand{\Delta}{\mathnormal{\oldDelta}}
\renewcommand{\Theta}{\mathnormal{\oldTheta}}
\renewcommand{\Pi}{\mathnormal{\oldPi}}
\renewcommand{\Upsilon}{\mathnormal{\oldUpsilon}}

\newcommand{\NN}{\mathcal{N}}
\newcommand{\FF}{\mathcal{F}}

\renewcommand{\AA}{\mathcal{A}}
\newcommand{\BB}{\mathcal{B}}
\newcommand{\MM}{{\mathcal M}}

\newcommand{\VV}{\mathcal{V}}
\newcommand{\PP}{\mathcal{P}}
\newcommand{\tr}{\mathop{\mathrm{tr}}}

\ifx\genfrac\sdflkaj

\else

\fi
\newcommand{\sfrac}[2]{{\textstyle\frac{#1}{#2}}}
\newcommand{\half}{\sfrac{1}{2}}

\newcommand{\quarter}{\sfrac{1}{4}}
\newcommand{\quart}{\sfrac{1}{4}}

\newcommand{\ut}{\tilde{u}}

\newcommand{\la}{\lambda}

\newcommand{\bJ}{\bar J}
\newcommand{\bI}{\bar{I}}
\newcommand{\bK}{\bar{K}}
\newcommand{\bL}{\bar{L}}
\newcommand{\btau}{{\bar \tau}}
\newcommand{\ba}{{\bar a}}
\newcommand{\bb}{{\bar b}}

\newcommand{\al}{\alpha}
\newcommand{\de}{\delta}

\newcommand{\nn}{\nonumber}

\makeatletter
\def\mr@ignsp#1 {\ifx\:#1\@empty\else #1\expandafter\mr@ignsp\fi}%
\newcommand{\multiref}[1]{\begingroup%\let\protect\string%
\xdef\mr@no@sparg{\expandafter\mr@ignsp#1 \: }%
\def\mr@comma{}%
\@for\mr@refs:=\mr@no@sparg\do{\mr@comma\def\mr@comma{,}\ref{\mr@refs}}%
\endgroup}
\makeatother

\newcommand{\hypref}[2]{\ifx\href\asklfhas #2\else\href{#1}{#2}\fi}

\renewcommand{\eqref}[1]{(\multiref{#1})}

\ifx\href\asklfhas\newcommand{\href}[2]{#2}\fi

\newcommand{\g}{\gamma}

\newcommand{\be}{\begin{eqnarray}}
\newcommand{\ee}{\end{eqnarray}}

\makeatletter \@addtoreset{equation}{section} \makeatother

\makeatletter
\let\old@startsection=\@startsection
\let\oldl@section=\l@section
\renewcommand{\@startsection}[6]{\old@startsection{#1}{#2}{#3}{#4}{#5}{#6\mathversion{bold}}}
\renewcommand{\l@section}[2]{\oldl@section{\mathversion{bold}#1}{#2}}
\makeatother

\makeatletter
\let\old@makecaption=\@makecaption
\def\@makecaption{\small\old@makecaption}
\makeatother

\renewcommand{\geq}{\geqslant}

\begin{document}

\thispagestyle{empty}
\begin{flushright}\footnotesize
\texttt{ITEP-TH-01/09}\\ \texttt{LPTENS-09/01}\\
\texttt{UUITP-01/09} \vspace{0.8cm}
\end{flushright}

\renewcommand{\thefootnote}{\fnsymbol{footnote}}
\setcounter{footnote}{0}

\begin{center}
{\Large\textbf{\mathversion{bold} Two loop integrability for Chern-Simons   \\
theories with ${\cal N}=6$ supersymmetry}\par}

\vspace{1.5cm}

\textrm{J.~A.~Minahan$^{1}$, W.~Schulgin$^{2,3}$ and
K.~Zarembo$^{2,1}$\footnote{Also at ITEP, Moscow, Russia}}
\vspace{8mm}

\textit{$^{1}$ Department of Physics and Astronomy, Uppsala University\\
SE-751 08 Uppsala, Sweden}\\
\texttt{joseph.minahan@fysast.uu.se} \vspace{3mm}

\textit{$^{2}$ Laboratoire de Physique Th\'eorique\footnote{Unit\'e
Mixte de Recherche du CNRS UMR 8549},
\'Ecole Normale Sup\'erieure\\
24 rue Lhomond, 75231 Paris, France }\\
\texttt{Waldemar.Schulgin@lpt.ens.fr,Konstantin.Zarembo@lpt.ens.fr}
\vspace{3mm}

\textit{$^{3}$Laboratoire de Physique Th\'eorique et Hautes
Energies\footnote{Unit\'e Mixte de Recherche du CNRS UMR 7589},\\
UPMC Univ Paris 06, Boite 126, 4 place Jussieu,\\ F-75252 Paris
Cedex 05 France}\\

%%%%%%%%

\par\vspace{1cm}

\textbf{Abstract} \vspace{5mm}

\begin{minipage}{14cm}

We consider two-loop anomalous dimensions for fermionic operators in
the ABJM model and the ABJ model.  We find the appropriate
Hamiltonian and show that it is consistent with a previously
predicted Bethe ansatz for the ABJM model.
The difference between the ABJ and ABJM models is invisible at the
two-loop level due to cancelation of parity violating diagrams.
We then construct a Hamiltonian for the full two-loop $OSp(6|4)$
spin chain by first constructing the Hamiltonian for an $SL(2|1)$
subgroup, and then lifting to
$OSp(6|4)$.  We show that this Hamiltonian is  consistent with the
Hamiltonian found  for the fermionic operators.

\end{minipage}

\end{center}

\vspace{0.5cm}

%%%%%%%%%%%%%%%%%%%%%%%%%%%%%%%%%%%%%%%%%%%%%%%%%%%%%%%%%%%%%%%%%%%%%%%%%%%%%%%%

\newpage
\setcounter{page}{1}
\renewcommand{\thefootnote}{\arabic{footnote}}
\setcounter{footnote}{0}

%%%%%%%%%%%%%%%%%%%%%%%%%%%%%%%%%%%%%%%%%%%%%%%%%%%%%%%%%%%%%%%%%%%%%%%%%%%%%%%%
\section{Introduction}

The ABJM  model \cite{Aharony:2008ug} has opened up a new
avenue in which to explore integrability in the planar limit of gauge
theories in three dimensions.
This model is comprised of two $U(N)$ gauge groups having a
Chern-Simons action with levels $k$ and $-k$ respectively. The model
also contains scalars and fermions that live in the bifundamental
representations of $U(N)\times U(N)$.  If the six scalar and two
scalar-two fermion couplings are properly tuned
\cite{Aharony:2008ug,Benna:2008zy,Hosomichi:2008jb}, then the theory
has an ${\cal N}=6$ supersymmetry.  This then leads to an
$SO(6)\simeq SU(4)$ $R$-symmetry, with the $(N,\bar{N})$ scalars
transforming in the {\bf 4} while the $(N,\bar{N})$ fermions
transform in the {$\bf\bar 4$}, with an appropriate $R$-symmetry
assignment for the conjugates. This theory also has a CP invariance.
The Chern-Simons action changes sign under parity, but can also be
accompanied by an exchange of the gauge groups to restore the sign.

Chern-Simons-matter theories admit
a
planar limit when the
rank of the gauge group and the Chern-Simons level simultaneously
approach
infinity such that their ratio $\la=N/k$ remains finite. The
operator mixing problem can then be reformulated in terms of a
quantum spin chain whose interaction range grows with the order of
perturbation theory. Interesting effects in Chern-Simons-matter
theories start at two loops \cite{Chen:1992ee}, and
with one loop terms absent the leading order spin-chain Hamiltonian
involves three-site next-to-nearest-neighbor interactions
\cite{Gaiotto:2007qi}. Such spin chains are usually not integrable,
which is indeed the case for the $\mathcal{N}=1$ superconformal
Chern-Simons \cite{Gaiotto:2007qi}.
However, in \cite{Minahan:2008hf} it was shown that the scalar
sector of the $\mathcal{N}=6$ ABJM theory is integrable at two loops
(see also \cite{Bak:2008cp}).
The corresponding spin chain is made up of alternating sites of {\bf
4} and $\bf \bar{4}$ spins, and the Hamiltonian contains terms up to
next to nearest neighbor interactions.  The CP invariance of the
gauge theory results in a parity invariance for the spin chain.  The
Bethe ansatz for this spin-chain was given and was shown to lead to
correct results for the anomalous dimensions of scalar operators.  A
conjectured Bethe ansatz was also given for the full superconformal
group $OSp(6|4)$. This Bethe ansatz is consistent with results from
the $SU(2|2)$ sector which was also shown to be integrable
\cite{Gaiotto:2008cg}. Furthermore, these Bethe ans\"atze were
extended to an all-loop Bethe ansatz
\cite{Gromov:2008qe,Ahn:2008aa}.  Other interesting results related
to the integrability of the ABJM model have been found in
\cite{Arutyunov:2008if,Stefanski:2008ik,Gromov:2008bz, Nishioka:2008gz,Grignani:2008is,Grignani:2008te,Astolfi:2008ji,Chen:2008qq,Lee:2008ui,Shenderovich:2008bs,Ahn:2008hj,Rashkov:2008rm,Ryang:2008rc,Bombardelli:2008qd,Lukowski:2008eq,Ahn:2008tv,Ahn:2008wd,Jain:2008mt,Kristjansen:2008ib,Abbott:2008qd,Sundin:2008vt,Bak:2008xq,Lee:2008yq}.

In  another development, Aharony, Bergman and Jafferis (ABJ)
extended the ABJM analysis to include among other things an
asymmetric theory with gauge group $U(N)\times U(\hat{N})$
\cite{Aharony:2008gk}.  Such a theory is no longer CP invariant if
$N\ne \hat{N}$,
but has all the same continuous symmetries as the $N=\hat{N}$ ABJM
model \cite{Hosomichi:2008jb}.
One would then expect that the parity invariance of the spin chain
is broken. This does not necessarily mean that the integrability is
lost, there are integrable spin chains with broken parity
\cite{Bak:2008vd}.

Nevertheless, one can
give a general argument
that the ABJ model
should not be integrable.
 The argument is based on the dual description in
terms of type IIA string theory on an $AdS_4\times CP^3$ background
\cite{Aharony:2008ug,Aharony:2008gk}. The integrability of the spin
chain is reflected in the classical integrability of the string
sigma-model
\cite{Arutyunov:2008if,Stefanski:2008ik,Gromov:2008bz}\footnote{The
integrability is manifest in the coset formulation of the
sigma-model, which corresponds to the Green-Schwarz action with
partially fixed $\kappa $-symmetry. It is not known if the full
Green-Schwarz sigma-model \cite{Gomis:2008jt} is integrable.},  %footnote added in v2
which may, or may not be preserved at the quantum level. The different
ranks
 in the ABJ model  means that there are two 't Hooft couplings,
 $\la=N/k$ and $\hat{\la}=\hat{N}/k$,
 and their difference
 leads to a
theta-angle on the string world-sheet:
$\vartheta _{\rm w.s.}\sim \lambda -\hat{\lambda }$
\cite{Aharony:2008gk}, which is responsible for
world-sheet parity
violation,
 analogous to potential parity violation in the
spin chain. However, it is commonly believed that the theta-angle
and parity violation destroy integrability. We can draw an analogy
with the $O(3)$ non-linear sigma-model which is integrable at
$\vartheta =0$ and $\vartheta =\pi $, but presumably is not
integrable at generic $\vartheta $ \cite{Zamolodchikov:1992zr}. %fixed in v2
The
analogy with the $O(3)$ model is arguably far-fetched, the
supercoset sigma-model on $AdS_4\times CP^3$ is quite different. So,
our argument is only qualitative, and it would be very interesting
to check if the ABJ model is integrable or not for generic $\la$ and
$\hat{\la}$. At strong coupling, the dependence on the difference
$\lambda -\hat{\lambda }$ is very weak, since the world-sheet
instanton effects which are sensitive to the theta-angle are
exponentially suppressed. But at weak coupling, one has no reason to
think that that the anomalous dimensions will depend on the
geometric mean of the 't~Hooft couplings, $\lambda \hat{\lambda }$,
and not on their difference,  $\lambda -\hat{\lambda }$ which is the
measure of parity violation.

Nonetheless, somewhat surprisingly, it was shown that the parity
invariance and with it the integrability in the scalar sector is
still preserved at the two loop level
for generic $\lambda $ and $\hat{\lambda }$
\cite{Bak:2008vd}. In principle, two loop corrections could come
with factors of $\la^2$, $\hat{\la}^2$ or $\la\hat{\la}$.
But one can quickly see
that in the scalar sector, only diagrams with factors of
$\la\hat{\la}$ appear.  Hence the parity is unbroken and the
integrability is preserved, with all factors of $\la^2$ in the ABJM
spin chain replaced by $\la\hat{\la}$ in the more general ABJ case.

Going outside the scalar sector,  one does find diagrams that are
proportional to $\la^2$ and $\hat{\la}^2$, so it is possible that
only the scalar sector is integrable.
Moreover,
the integrability of the full ABJM model
is only a conjecture and has to be verified, especially in view of
the
discrepancies that have arisen between string theory results
\cite{McLoughlin:2008ms,Alday:2008ut,Krishnan:2008zs,McLoughlin:2008he}
and those derived from the algebraic curve
\cite{Gromov:2008bz,Gromov:2008qe,Gromov:2008fy}.

In this paper we will explore these  issues by first considering
gauge invariant operators with one fermion present.  Here we will
find that there are loop diagrams that lead to possible $\la^2$ and
$\hat{\la}^2$ terms.  However, these diagrams cancel off with other
diagrams, resulting in a dilatation operator with $\la\hat{\la}$
terms only.  We compute the dilation operator for all operators of
this type and explicitly show that it is consistent with the
proposed Bethe ansatz in \cite{Minahan:2008hf} for operators of
length 4.

We then go on to construct the general  Hamiltonian for the full
$OSp(6|4)$.  We do this by first constructing it for an
$SL(2|1)\simeq OSp(2|2)$ sector, a noncompact closed
sector containing fermions.  In fact the $R$-matrices for noncompact
$SL(2|1)$ have previously appeared in the literature
\cite{Derkachov:2000ne,Belitsky:2006cp}\footnote{Spin chains have
also been constructed for finite representations of $SL(2|1)$
\cite{Links,Abad,Essler:2005ag}}.  The Hamiltonian is written in
terms of projectors onto irreducible representations of $SL(2|1)$.
The results for $SL(2|1)$ can then be uniquely lifted to the full
$OSp(6|4)$ in a way analogous to how the $SU(1,1)$ sector of ${\cal
N}=4$ Super Yang-Mills is lifted to the full $PSU(2,2|4)$
superconformal group \cite{Beisert:2003jj,Beisert:2003yb}.  The
difference here is that for this spin chain the $R$-matrices are
built out of two complex representations, while in $SU(1,1)$ the
$R$-matrix is built using a single real representation.

In section 2 we review the Lagrangian of ${\cal N}=6$ superconformal Chern-Simons as well as the conjectured two-loop Bethe equations for all single trace operators.   In section 3 we derive the Hamiltonian for the spin chain with one fermion present by computing the appropriate Feynman diagrams.  In section 4 we compute the anomalous dimensions for all length 4 operators with one fermion and show that the conjectured Bethe equations give the correct result.  In section 5 we construct the Hamiltonian for the full $OSp(6|4)$.  We do this by taking the $R$-matrix for the $SL(2|1)$ subgroup and extending it to $OSp(6|4)$.  We then show that the resulting Hamiltonian is consistent with the Hamiltonian for scalar operators and the Hamiltonian derived in section 3 for operators with one fermion. In section 6 we give a short discussion.

{\it Note added:} As this paper was being prepared we received \cite{Zwiebel:2009vb} which overlaps with our results in section 5.
%%%%%%%%%%%%%%%%%%%%%%%%%%%%%%%%%%%%%%%%%%%%%%%%%%%%%%%%%%%%%%%%%%%%%%%%%%%%%%%%
\section{The ${\cal N}=6$ theory and its Bethe equations}

In this section we give a short  review of the ABJM ${\cal N}=6$
superconformal Chern-Simons theory and the conjectured two-loop
Bethe equations that arise from it.

The lagrangian of the $\mathcal{N}=6$ Chern-Simons theory is
\cite{Aharony:2008ug,Benna:2008zy,Hosomichi:2008jb}
\begin{eqnarray}\label{Lagrangian}
 \mathcal{L}&=&\frac{k}{4\pi }\,{\rm tr}\,\left[
 \varepsilon ^{\mu \nu \lambda }\left(
 A_\mu \partial _\nu A_\lambda +\frac{2}{3}\,A_\mu A_\nu A_\lambda
 -\hat{A}_\mu \partial _\nu \hat{A}_\lambda
 -\frac{2}{3}\,\hat{A}_\mu \hat{A}_\nu \hat{A}_\lambda
 \right)\right.
 \nonumber \\
 &&\left.
 +  D_\mu Y^\dagger _AD^\mu Y^A +i\bar{\psi }^AD\!\!\!\!/\,\psi_A
 +\frac{1}{12}\,Y^AY^\dagger _AY^BY^\dagger _BY^CY^\dagger _C
 +\frac{1}{12}\,Y^AY^\dagger _BY^BY^\dagger _CY^CY^\dagger _A
 \right.
 \nonumber \\
 &&\left.
 -\frac{1}{2}\,Y^AY^\dagger _AY^BY^\dagger _CY^CY^\dagger _B
 +\frac{1}{3}\,Y^AY^\dagger _BY^CY^\dagger _AY^BY^\dagger _C
   \right.
 \nonumber \\
 &&\left.
 -\frac{1}{2}\,Y^\dagger _AY^A\bar{\psi }^B\psi _B
 +Y^\dagger _AY^B\bar{\psi }^A\psi _B
 +\frac{1}{2}\,\bar{\psi }^AY^BY^\dagger _B\psi _A
 -\bar{\psi }^AY^BY^\dagger_A\psi _B
  \right.
 \nonumber \\
 &&\left.
 +\frac{1}{2}\,\varepsilon ^{ABCD}Y^\dagger _A\bar{\psi} _{cB}Y^\dagger _C\psi _D
 -\frac{1}{2}\,\varepsilon _{ABCD}Y^A\bar{\psi }^BY^CC\psi_c ^D
 \right ],
\end{eqnarray}
where $D_\mu X=\partial X+A_\mu X-X\hat{A}_\mu $, $D_\mu X^\dagger
=\partial _\mu X^\dagger +\hat{A}_\mu X^\dagger -X^\dagger A_\mu $.
We use the $+--$ conventions for the metric. Raised capital latin
indices (A,B,C,D) are $SU(4)$ $R$-symmetry indices for the
fundamental representation, while lowered indices of this type are
for the anti-fundamental representation.  In the last two terms,
$\psi_c^A$ is the charge conjugated fermion field,
$\psi_c^A=C\gamma_0\,\psi_A^*$, where $C$ is the charge conjugation
matrix: $\gamma _\mu C+C\gamma _\mu ^T=0$, $C^2=1$, $C^\dagger =C$,
and $C^T=-C$. The Dirac matrices satisfy
\begin{equation}\label{}
 \gamma ^\mu \gamma ^\nu =\eta ^{\mu \nu }+i\varepsilon ^{\mu \nu \lambda }\gamma
 _\lambda ,
\end{equation}
with $\varepsilon ^{012}=1=\varepsilon _{012}$.  The form of the
Lagrangian is the same for the ABJ model where the number of colors
for the two gauge groups, $N$ and $\hat{N}$, are no longer equal.

The ABJM model is conformally invariant. Its Hilbert space consists
of all possible local operators, which at tree level are highly
degenerate. The degeneracy is lifted by quantum corrections and leads
to a fairly complicated mixing pattern. The mixing matrix can be
calculated in perturbation theory, and in the large-$N$ limit can be
identified with the Hamiltonian of a quantum spin chain. Computation
of the full mixing matrix is a formidable task even at the leading
two-loop order, and we will first consider closed sectors for which
the Hamiltonian can be evaluated directly from Feynman diagrams with
relative ease. One such sector is comprised of the bosonic
operators\footnote{These operators form a closed sector only at two
loops.}:
\begin{equation}\label{bosonicops}
 \tr Y^{A_1}Y^\dagger_{B_1}Y^{A_2}Y^\dagger_{B_2}\dots
 Y^{A_L}Y^\dagger_{B_L},
\end{equation}
which form a basis of states in the closed alternating spin chain
with $SU(4)$ symmetry. The
spin chain Hamiltonian is \cite{Minahan:2008hf,Bak:2008vd}
 \begin{equation}\label{dilopbos}
 \Gamma_{{\rm bos}} =\frac{\lambda \hat{\lambda }}{2}\sum_{l=1}^{2L}\left(
 2-2P_{l,l+2}+P_{l,l+2}K_{l,l+1}+K_{l,l+1}P_{l,l+2}
 \right),
\end{equation}
where $K_{ll'}$ and $P_{ll'}$ are the trace and permutation
operators acting on the $l$th and $l'$th sites of the lattice:
$P^{AA'}{}_{BB'}=\delta ^A{}_{B'}\,\delta ^{A'}{}_{B}$,
$K^A_{A'}\,{}^{B'}_{B}=\delta ^A{}_{B}\,\,\delta _{A'}{}^{B'}$.

In \cite{Minahan:2008hf} a set of Bethe  equations was given that
are consistent with the Hamiltonian in (\ref{dilopbos}).  Based on
the $OSp(6|4)$ superconformal algebra, an extended set of Bethe
equations were conjectured valid for all operators.  These equations
are
\begin{figure}[t]
\centerline{\includegraphics[width=8cm]{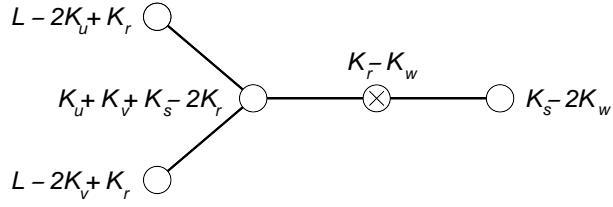}}
\caption{\label{dynya}\small The Dynkin diagram of $OSp(6|4)$. Shown
are the Dynkin labels of the state with $\{K_u,K_v,K_r,K_s,K_w\}$
Bethe roots.}
\end{figure}
\begin{eqnarray}
 \label{betheeqscf}
\left(\frac{u_j+i/2}{u_j-i/2}\right)^L&=&\prod_{k=1,k\ne j}^{K_u}\frac{u_j-u_k+i}{u_j-u_k-i} \prod_{k=1}^{K_r}\frac{u_j-r_k-i/2}{u_j-r_k+i/2}\nn\\
\left(\frac{v_j+i/2}{v_j-i/2}\right)^L&=&\prod_{k=1,k\ne
j}^{K_v}\frac{v_j-v_k+i}{v_j-v_k-i}
\prod_{k=1}^{K_r}\frac{v_j-r_k-i/2}{v_j-r_k+i/2}\nn\\
1&=& \prod_{k=1,k\ne j}^{K_r}\frac{r_j-r_k+i}{r_j-r_k-i}\prod_{k=1}^{K_u}\frac{r_j-u_k-i/2}{r_j-u_k+i/2}\prod_{k=1}^{K_v}\frac{r_j-v_k-i/2}{r_j-v_k+i/2}\prod_{k=1}^{K_s}\frac{r_j-s_k-i/2}{r_j-s_k+i/2}\nn\\
1&=&\prod_{k=1}^{K_r}\frac{s_j-r_k-i/2}{s_j-r_k+i/2}\prod_{k=1}^{K_w}\frac{s_j-w_k+i/2}{s_j-w_k-i/2}\nn\\
1&=&\prod_{k=1,k\ne
j}^{K_w}\frac{w_j-w_k-i}{w_j-w_k+i}\prod_{k=1}^{K_s}\frac{w_j-s_k+i/2}{w_j-s_k-i/2}\,.
\end{eqnarray}
The first three lines in (\ref{betheeqscf}) are the Bethe equations
of scalar operators when $K_s=0$. The Bethe equations correspond to
the Dynkin diagram in fig.~\ref{dynya}. The excitation numbers $K_a$
must satisfy
a  set of inequalities (which are basically the highest-weight
conditions for the Bethe wave functions):
\begin{eqnarray}\label{condsonweights}
 L-2K_u+K_r&\geq&0 \nonumber \\
 L-2K_v+K_r&\geq&0 \nonumber \\
 K_u+K_v-2K_r+1&\geq&0 \nonumber \\
 K_r&>&K_s \nonumber \\
 K_s-2K_w&\geq&0.
\end{eqnarray}

The solutions that correspond to gauge-theory operators in addition
satisfy the level-matching (zero-momentum) condition,
\begin{equation}\label{mmcond}
 \prod_{j=1}^{K_u}\frac{u_j+i/2}{u_j-i/2}\prod_{j=1}^{K_v}\frac{v_j+i/2}{v_j-i/2}=1.
\end{equation}
The  anomalous dimensions for the operators
are eigenvalues of
the Hamiltonian for the spin chain, and
are given by
\begin{equation}\label{betheen}
 \gamma =\lambda \hat{\lambda }\left(\sum_{j=1}^{K_u}\frac{1}{u_j^2+
 \frac{1}{4}}+\sum_{j=1}^{K_v}\frac{1}{v_j^2+\frac{1}{4}}\right).
\end{equation}
Notice that only the $u_j$ and $v_j$ Bethe roots contribute  to
(\ref{mmcond}) and (\ref{betheen}).
Only these roots
carry momentum and energy.
%%%%%%%%%%%%%%%%%%%%%%%%%%%%%%%%%%%%%%%%%%%%%%%%%%%%%%%%%%%%%%%%%%%%%%%%%%%%%%%%
\section{Fermionic part of the dilatation operator}

In this section  we study operators with a single fermion insertion,
\begin{equation}\label{}
\tr\psi _{B}Y^\dagger_{B_1}Y^{A_2}Y^\dagger_{B_2}\dots
 Y^{A_L}Y^\dagger_{B_L}\qquad {\rm and}\qquad
\tr Y^{A_1}\psi_c^{A}Y^{A_2}Y^\dagger_{B_2}\dots
 Y^{A_L}Y^\dagger_{B_L}.
\end{equation}
They also form a closed sector at two loops because of fermion
number conservation.
We will compute the two-loop mixing matrix for these operators.
In the bulk of the operator (far from the fermion insertion),
the mixing matrix
is the same as in (\ref{dilopbos}). Here we concentrate on the
mixing that involves the fermion insertion.

\begin{figure}[t]
\centerline{\includegraphics[height=2.5cm]{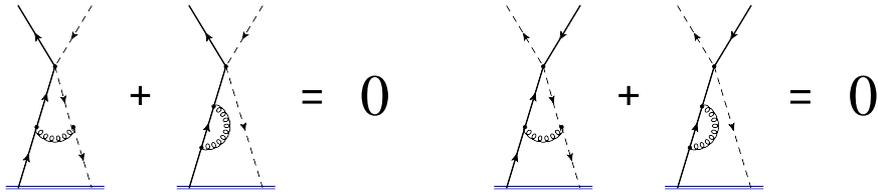}}
\caption{\label{canc}\small The CP-breaking contributions to the
dilatation operator mutually cancel.}
\end{figure}

Potentially, mixing of the fermions could depend on $\lambda ^2$ and
$\hat{\lambda }^2$ separately and thus could break CP invariance.
The $\lambda $--$\hat{\lambda }$ power counting can be easily
pictured by coloring the double-line 't~Hooft diagrams. There are
two gauge groups, and one can draw the $U(N)$ and $U(\hat{N})$ index
lines in two different colors, say blue and red. A planar diagram
will then be a collection of facets (index loops) painted in the two
colors. Since the matter fields are bifundamentals and the gauge
fields are adjoints, the color changes across a scalar or fermion
propagator, but stays the same across a gauge or ghost propagator. A
diagram with $F_r$ red facets and $F_b$ blue facets is accompanied
by a factor $N^{F_r}\hat{N}^{F_b}\propto\lambda ^{F_r}\hat{\lambda
}^{F_b}$. It is the diagrams with $F_r\neq F_b$ %fixed in v2
that potentially
violate CP. At the two-loop order there are four such diagrams, all
of which involve internal gauge-boson lines. To our surprise these
diagrams mutually cancel (fig.~\ref{canc}). The fermion part of the
dilatation operator thus is also proportional to $\lambda
\hat{\lambda }$:
\begin{equation}\label{dil1}
 \Gamma _{\rm ferm}=\frac{\lambda \hat{\lambda }}{6}
 ({\rm const}+H),
\end{equation}

\begin{figure}[t]
\centerline{\includegraphics[height=2.5cm]{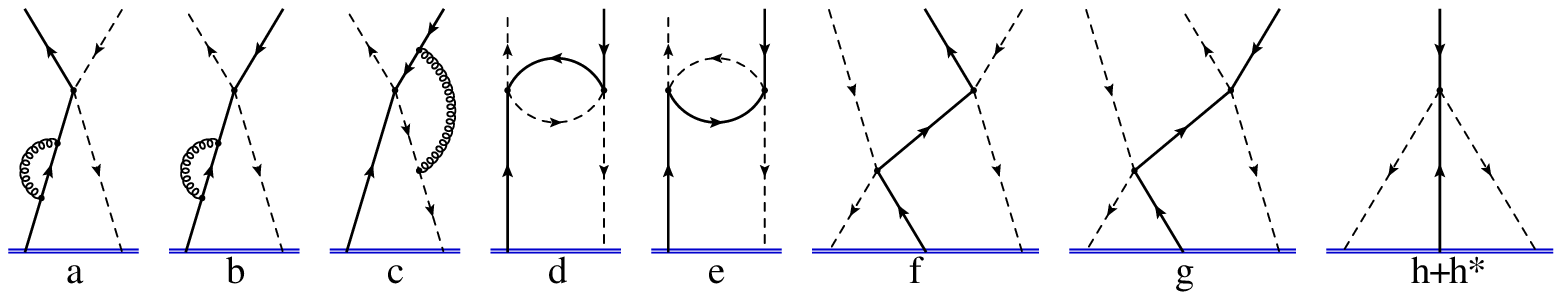}}
\centerline{\includegraphics[height=2.5cm]{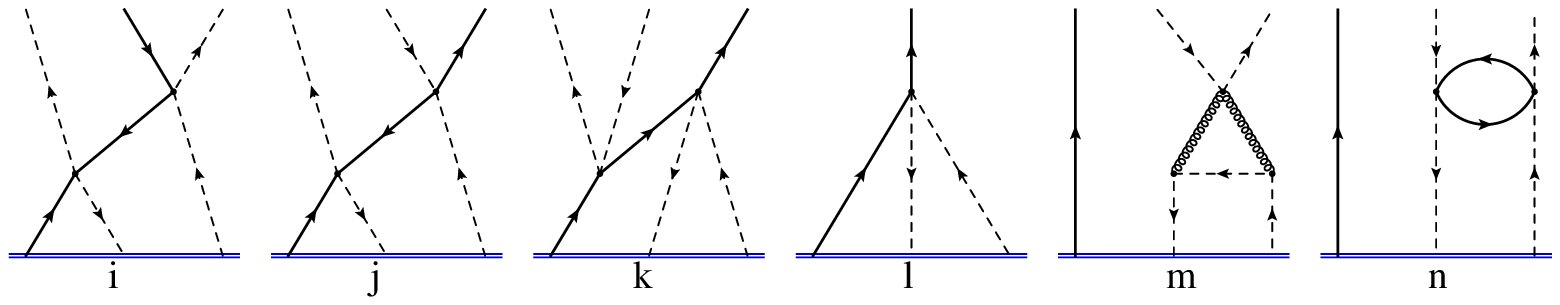}}
\caption{\label{omix}\small The diagrams that contribute to the
mixing of operators with single fermion insertion.}
\end{figure}

 The
diagrams that contribute to the fermion  mixing and that do not
identically vanish are listed in fig~\ref{omix}. There is also a
number of diagrams that do not mix different operators and only
contribute to the constant term in (\ref{dil1}). We have not shown
these diagrams and have not computed them. The constant will be
later fixed by requiring that the dilatation operator preserves
supersymmetry. Each diagram in fig.~\ref{omix} should be
supplemented with its parity conjugate (denoted by a *),
The only exception is
he diagram $h$ which is its own parity-conjugate.
The last two diagrams, $m$ and $n$, do not directly involve fermion
interaction vertices, and one may be tempted to attribute them to
the bosonic part of the mixing matrix (\ref{dilopbos}). Indeed,
these diagrams combine with the six-boson vertex to cancel off
the nearest-neighbor exchanges in the bosonic part of the mixing
matrix \cite{Minahan:2008hf}. In the presence of the fermion
insertion, this cancelation is incomplete. In the middle of the
scalar operator, $m$ and $n$ participate in the cancelation of the
six-vertex graph on the left and on the right, but now there is a
fermion on the left, and half of the $m+n$ has nothing to cancel.
The easiest way to take into account these diagrams is to add the
nearest-neighbor term from the six-vertex graph (which can be found
in eq.~(2.6) of \cite{Minahan:2008hf}) to the fermion mixing matrix
with coefficient $(-1/2)$. The diagrams $h+h^*$ and $l$ represent
mixing with a derivative of $\psi_c$ and $\psi$. However, because of
the Lorentz invariance the derivative must be accompanied by the
Dirac matrix and combined into $D\!\!\!\!/\ \psi$ or $D\!\!\!\!/\
\psi_c$,
which can be eliminated by use of the
equations of motion.

Computing the diagrams in fig.~\ref{omix} is in principle
straightforward.
Collecting all pieces together, we get:
\begin{eqnarray}\label{h3}
 H\circ Y^\dagger _A\psi _BY^\dagger _C&=&
 -2Y^\dagger _A\psi _CY^\dagger _B
 -2Y^\dagger _B\psi _AY^\dagger _C
 -Y^\dagger _B\psi _CY^\dagger _A
 -Y^\dagger _C\psi _AY^\dagger _B
 -2Y^\dagger _C\psi _BY^\dagger _A
 \nonumber \\ &&
 -6\varepsilon _{BCDE}Y^\dagger _AY^D \psi_c^E
 -6\varepsilon _{ABDE} \psi_c^DY^EY^\dagger _C
 -\varepsilon _{ABCD} \psi_c^DY^EY^\dagger _E
 \nonumber \\ &&
 +3\varepsilon _{ACDE}Y^\dagger _BY^D\psi _c^E
 +3\varepsilon _{ACDE}\psi _c^DY^EY^\dagger _B
 +\varepsilon _{ABCD}Y^\dagger _EY^E\psi_c^D
  \nonumber \\ &&
 -2\varepsilon _{ABCD}Y^\dagger _EY^D\psi_c^E
 +2\varepsilon _{ABCD}\psi _c^EY^DY^\dagger _E,
\end{eqnarray}
\begin{eqnarray}\label{h4}
 H\circ \psi _AY^\dagger _BY^C&=&
 -3Y^CY^\dagger _B\psi _A
 -3Y^CY^\dagger _A\psi _B \nonumber \\ &&
 +\delta _B^C Y^DY^\dagger _A\psi _D
 +\delta _B^CY^DY^\dagger _D\psi _A
 +\delta _A^C Y^DY^\dagger _D\psi _B
 +\delta _A^CY^DY^\dagger _B\psi _D \nonumber \\ &&
 +2\delta _B^C\psi _AY^\dagger _DY^D
 -\delta _B^C\psi _DY^\dagger _AY^D
 -\delta _A^C\psi _BY^\dagger _DY^D
 +2\delta _A^C\psi _DY^\dagger _BY^D \nonumber \\ &&
 +3\varepsilon _{ABDE}Y^D\psi_c^CY^E
 +\delta _B^C\varepsilon _{ADEF}Y^D\psi_c^EY^F
 -2\delta _A^C\varepsilon _{BDEF}Y^D\psi_c^EY^F,\nonumber \\ &&
\end{eqnarray}
\begin{eqnarray}\label{h5}
 H\circ Y^AY^\dagger _B\psi _C&=&
 -3\psi _CY^\dagger _BY^A
 -3\psi _BY^\dagger _CY^A \nonumber \\ &&
 +\delta _B^A\psi _DY^\dagger _CY^D
 +\delta _B^A\psi _CY^\dagger _DY^D
 +\delta _C^A\psi _BY^\dagger _DY^D
 +\delta _C^A\psi _DY^\dagger _BY^D \nonumber \\ &&
 +2\delta _B^AY^DY^\dagger _D\psi _C
 -\delta _B^AY^DY^\dagger _C\psi _D
 -\delta _C^AY^DY^\dagger _D\psi _B
 +2\delta _C^AY^DY^\dagger _B\psi _D
 \nonumber \\ &&
 +3\varepsilon _{BCDE}Y^D\psi_c^AY^E
 -\delta _B^A\varepsilon _{CDEF}Y^D\psi_c^EY^F
 +2\delta _C^A\varepsilon _{BDEF}Y^D\psi_c^EY^F.\nonumber \\ &&
\end{eqnarray}
There are two terms in the Lagrangian (\ref{Lagrangian})
contributing to
the $ Y^\dagger Y \bar{\psi} \psi$-vertex and two terms contributing
to the $ \bar{\psi} Y Y^\dagger \psi$-vertex.
The terms $
\,{\rm tr}\,Y^\dagger_A Y^B \bar{\psi}^A\psi_B$ and $
\,{\rm tr}\,\bar{\psi}^A Y^B Y_A^\dagger \psi_B$
mix the flavor indices while the other two do not. It means the
diagrams with two fermion-boson vertices correspond to up to four
different terms in the Hamiltonian.

To get the action of the dilatation operator on the states with a
$\psi_c$ insertion one has to compute the same set of diagrams with
the arrows on all the lines inverted. However, pulling the charge
conjugation matrix through the fermion line reverses all the momenta
due to the identity $-C/\!\!\!{p}^T=/\!\!\!{p}C$, so the result will
be exactly the same.  Consequently, the dilatation operator is given
by interchanging the upper and lower indices in
(\ref{h3})--(\ref{h5}),
 and so making the replacement
$Y^A\leftrightarrow Y^\dagger _A$, $\psi_A\leftrightarrow \psi_c^A$,
$\varepsilon _{ABCD}\leftrightarrow\varepsilon ^{ABCD}$, $\delta
_A^B\leftrightarrow\delta _B^A$.

The constant in (\ref{dil1}) can be fixed by requiring that the
ground state energy is zero. The computation at length-$4$
(sec.~\ref{length4}) gives:
\begin{equation}\label{dil}
 \Gamma _{\rm ferm}=\frac{\lambda \hat{\lambda }}{6}
 \left(20+H\right).
\end{equation}
The full one-loop dilatation operator is then the sum of the mixing
matrices acting on scalars, eq.~(\ref{dilopbos}), and the fermion
mixing matrix (\ref{dil}),
\begin{equation}\label{}
 \Gamma =\Gamma _{\rm bos}+\Gamma _{\rm ferm}.
\end{equation}

\section{Length-$4$ operators}\label{length4}

In this section we will explicitly diagonalize the mixing matrix
from the previous section for operators of length four. We will also
solve the Bethe equations (\ref{betheeqscf}) for $L=2$ and compare
the resulting spectra of anomalous dimensions. There are in total
$2\times 4^4=512$ operators, but many of them are super-descendants
of the bosonic length-$4$ operators $\,{\rm tr}\,(YY^\dagger)^2 $.
The solutions of the Bethe equations describe primary operators, so
first we will discuss constraints imposed on the spectrum by
supersymmetry.

Under the $SU(4)$ R-symmetry, the length-$4$ states $\,{\rm
tr}\,\psi Y^\dagger YY^\dagger $ and $\,{\rm tr}\,Y\psi _cYY^\dagger
$ transform as
\begin{eqnarray}\label{l4_1}
 \bar{\mathbf{4}}\otimes\mathbf{4}\otimes\mathbf{4}\otimes\mathbf{4}&=&
 \mathbf{6}^3\oplus\mathbf{10}^3\oplus\bar{\mathbf{10}}\oplus\mathbf{64}^2\oplus\mathbf{70}
 \\  \label{l4_2}
 \mathbf{4}\otimes\bar{\mathbf{4}}\otimes\bar{\mathbf{4}}\otimes\bar{\mathbf{4}}&=&
 \mathbf{6}^3\oplus\mathbf{10}\oplus\bar{\mathbf{10}}^3\oplus\mathbf{64}^2\oplus\bar{\mathbf{70}}.
\end{eqnarray}
The scalar operators $\,{\rm tr}\,(YY^\dagger)^2 $ are in
\begin{equation}\label{}
 \bar{\mathbf{4}}\otimes{\mathbf{4}}\otimes\bar{\mathbf{4}}\otimes{\mathbf{4}}
 =\mathbf{1}^2\oplus\mathbf{15}^2\oplus\mathbf{20}\oplus\mathbf{84}.
\end{equation}
Their anomalous dimensions, in the units of $\lambda \hat{\lambda
}$, are $2,10,6,6,8,0$, respectively \cite{Minahan:2008hf}.

The supercharges act on the scalars as
\begin{eqnarray}\label{susy}
 Q_{AB}Y^C=\delta^C_A\psi _B-\delta _B^C\psi _A\\
 \label{susyprim}
 Q_{AB}Y^\dagger _C=\varepsilon _{ABCD}{\psi }_c^D.
\end{eqnarray}
Consequently,
$$
 Q\,{\rm tr}\,YY^\dagger YY^\dagger \sim \psi Y^\dagger YY^\dagger
 +Y{\psi }_cYY^\dagger .
$$
Since the supercharges are in the $\mathbf{6}$ of $SU(4)$, the
superpartners of the scalar operators belong to
\begin{equation}\label{}
 \mathbf{6}\otimes\left(\mathbf{1}^2\oplus\mathbf{15}^2\oplus\mathbf{20}\oplus\mathbf{84}\right)
 = \mathbf{6}^5\oplus\mathbf{10}^2\oplus\bar{\mathbf{10}}^2\oplus\mathbf{64}^4
 \oplus\mathbf{70}\oplus\bar{\mathbf{70}}{\color{gray}\oplus\mathbf{50}\oplus\mathbf{300}}.
\end{equation}
One has to remember, however, that not all representations in the
product are associated with operators. The last two representations
shown in gray are projected out. This can be understood from the
supersymmetry transformations (\ref{susy}), (\ref{susyprim}). The
left-hand side of (\ref{susy}) is in
$\mathbf{6}\otimes\mathbf{4}=\bar{\mathbf{4}}{\color{gray}\oplus\mathbf{20}}$,
but only the $\bar{\mathbf{4}}$ appears on the right-hand side.
Likewise, in (\ref{susyprim}), which is
$\mathbf{6}\otimes\bar{\mathbf{4}}=\mathbf{4}{\color{gray}\oplus\bar{\mathbf{20}}}$,
the $\bar{\mathbf{20}}$ is projected out.

The supersymmetry fixes part of the spectrum in (\ref{l4_1}),
(\ref{l4_2}):
\begin{eqnarray}\label{descendants}
 \mathbf{64}
 \oplus\mathbf{70}\oplus\bar{\mathbf{70}}: && 0\nonumber \\
 \mathbf{6}:&& 2\lambda \hat{\lambda }\nonumber \\
 \mathbf{6}^2\oplus\mathbf{10}^2\oplus\bar{\mathbf{10}}^2\oplus\mathbf{64}^2:
 && 6\lambda \hat{\lambda }\nonumber \\
 \mathbf{6}\oplus\mathbf{64}:&& 8\lambda \hat{\lambda }\nonumber \\
 \mathbf{6}:&& 10\lambda \hat{\lambda },
\end{eqnarray}
Hence, we are left with five multiplets whose highest-weight states
are fermionic length-$4$ operators:
\begin{equation}\label{hweights}
 \mathbf{6}\oplus\mathbf{10}^2\oplus\bar{\mathbf{10}}^2.
\end{equation}
We will compute their anomalous dimensions first by diagonalizing
the Hamiltonian (\ref{h3})--(\ref{h5}), (\ref{dilopbos}), and then
by solving the Bethe equations. We should note that the solutions of
the Bethe equations for the fermion states are sensitive to the
whole structure of the Dynkin diagram.

\subsection{Spectrum from the mixing matrix}

Let us diagonalize the Hamiltonian in (\ref{dil}). The simplest case
is the $\bar{\mathbf{70}}$ operator:
\begin{equation}\label{}
 O_{\bar{\mathbf{70}}}=\,{\rm tr}\,Y^\dagger _{(A}\psi _BY^\dagger
 _{C)}Y^D-{\rm traces}.
\end{equation}
The symmetrization acts on all three lower indices. The bosonic part
of the dilatation operator annihilates $O_{\bar{\mathbf{70}}}$,
because of the symmetry in $A$, $B$ and $C$. This operator does not
mix with the $\psi_c$ states either, since such mixing inevitably
involves contraction with the $\varepsilon _{ABFE}$ tensor. The rest
of the Hamiltonian in (\ref{h3})--(\ref{h5}) acts as
\begin{equation}\label{}
 H\circ
 O_{\bar{\mathbf{70}}}=-20\,O_{\bar{\mathbf{70}}}\,.
\end{equation}
The $\bar{\mathbf{70}}$ is a part of the BPS supermultiplet and thus
should have zero anomalous dimension. This fixes the constant term
in (\ref{dil}).

The next set of operators are the four states in the
$\bar{\mathbf{10}}$:
\begin{eqnarray}\label{}
 &&\mathcal{O}^{\bar{\mathbf{10}}}_{1\,AB}=\,{\rm tr}\,Y^\dagger _{C}\psi
 _{(A}Y^\dagger _{B)}Y^C,\qquad
 \mathcal{O}^{\bar{\mathbf{10}}}_{2\,AB}=\,{\rm tr}\,Y^\dagger _{(A}\psi
 _{B)}Y^\dagger _{C}Y^C,
 \nonumber \\
 &&\mathcal{O}^{\bar{\mathbf{10}}}_{3\,AB}=\,{\rm tr}\,Y^\dagger _{(A}\psi
 _{C}Y^\dagger _{B)}Y^C,\qquad
 \mathcal{O}^{\bar{\mathbf{10}}}_{4\,AB}=\,{\rm tr}\,Y^{C}{\psi
 }_c^{D}Y^{E}Y^\dagger _{(A}\varepsilon _{B)CDE}.
\end{eqnarray}
Here we also know part of the spectrum from supersymmetry,
eq.~(\ref{descendants}), but in addition to the two descendants
there are two highest-weight states. It turns out that the
dilatation operator is fully degenerate in this sector:
\begin{equation}\label{}
 \left.\Gamma\right|_{\mathbf{10}} =\lambda \hat{\lambda }
 \begin{pmatrix}
  6 & 0 & 0  & 0 \\
   0 & 6 & 0 & 0  \\
    0 & 0 & 6  & 0 \\
   0 & 0 & 0  & 6  \\
 \end{pmatrix}
\end{equation}
This agrees with (\ref{descendants}), and predicts that four of the
anomalous dimensions in (\ref{hweights}) are equal to $6\lambda
\hat{\lambda }$ (the anomalous dimensions in $\mathbf{10}$ are
obviously the same as in $\bar{\mathbf{10}}$).

The basis of operators in the $\mathbf{6}$ is spanned by
\begin{eqnarray}\label{}
 &&\mathcal{O}^{\mathbf{6}}_{1\,AB}=2\,{\rm tr}\,Y^\dagger _{C}\psi
 _{[A}Y^\dagger _{B]}Y^C,\qquad
 \mathcal{O}^{{\mathbf{6}}}_{2\,AB}=2\,{\rm tr}\,Y^\dagger _{[A}\psi
 _{B]}Y^\dagger _{C}Y^C,
 \nonumber \\
 &&\mathcal{O}^{{\mathbf{6}}}_{3\,AB}=2\,{\rm tr}\,Y^\dagger _{[A}\psi
 _{C}Y^\dagger _{B]}Y^C,\qquad
 \mathcal{O}^{\rm \mathbf{6}}_{4\,AB}=\varepsilon _{ABDE}\,{\rm
 tr}\,Y^C{\psi }_c^DY^EY^\dagger _C,
 \nonumber \\
 &&\mathcal{O}^{\rm \mathbf{6}}_{5\,AB}=\varepsilon _{ABDE}\,{\rm
 tr}\,Y^D{\psi }_c^EY^CY^\dagger _C,\qquad
 \mathcal{O}^{\rm \mathbf{6}}_{6\,AB}=\varepsilon _{ABDE}\,{\rm
 tr}\,Y^D{\psi }_c^CY^EY^\dagger _C.
\end{eqnarray}
In this basis,
\begin{equation}\label{}
\left.\Gamma\right|_{\mathbf{6}} =\frac{\lambda \hat{\lambda }}{3}
 \begin{pmatrix}
  {19} & -{5} & -{1}  & {7} & {1} & -{1}  \\
  -{5} & {19} & -{1}  & {1} & {7} & -{1} \\
  -{4} & -{4} & {22}  & {2} & {2} & -{2} \\
  {7} & {1} & -{1}  & {19} & -{5} & -{1} \\
  {1} & {7} & -{1}  & -{5} & {19} & -{1} \\
  {2} & {2} & -{2}  & -{4} & -{4} & {22} \\
 \end{pmatrix}
\end{equation}
This matrix has eigenvalues
\begin{equation}\label{}
 \left.\gamma\right|_{\mathbf{6}}=\lambda \hat{\lambda }\{10,8,8,6,6,2\},
\end{equation}
in agreement with (\ref{descendants}). The anomalous dimension of
the unique highest-weight $\mathbf{6}$ operator is equal to
$8\lambda \hat{\lambda }$.

\subsection{Spectrum from the Bethe equations}

The states with one fermion impurity correspond to the solutions of
the Bethe equations (\ref{betheeqscf}) with $K_s=1$. For $L=2$
(length-$4$ operators), the highest-weight conditions admit three
possible configurations of the Bethe roots, see diagram~\ref{dynya}
and eqs.~(\ref{condsonweights}): (i) $K_u=2$, $K_v=1$, $K_r=2$; (ii)
$K_u=1$, $K_v=2$, $K_r=2$; and (iii) $K_u=2$, $K_v=2$, $K_r=2$. We
can read off their $SO(6)$ quantum numbers from fig.~\ref{dynya}:
(i) is the $\bar{\mathbf{10}}$ (the Dynkin labels  are $[0,0,2])$;
(ii) is the $\mathbf{10}$ with the $[2,0,0]$ Dynkin labels; and
(iii) is $\mathbf{6}$ with $[0,1,0]$.

In all three cases, the Bethe equations simplify and reduce to
quadratic equations. For the case (i), there are two inequivalent
solutions that satisfy the momentum condition
(\ref{mmcond})\footnote{There is also an extra solution which has
non-zero momentum.}:
\begin{eqnarray}\label{}
 &&u_1=\pm\frac{\sqrt{11}+\sqrt{3}}{8}\,,\qquad %fixed in v2
 u_2=\mp\frac{\sqrt{11}-\sqrt{3}}{8}\,,\qquad
 v_1=\pm\frac{\sqrt{3}}{2}\,\qquad
 \nonumber \\ &&r_1=0,\qquad
 r_2=\pm\frac{\sqrt{3}}{2}\,,\qquad
 s_1=\pm\frac{\sqrt{3}}{4}\,.
\end{eqnarray}
They form a parity pair \cite{Beisert:2003tq}, and are degenerate in
energy:
\begin{equation}\label{}
 \gamma _{\bar{\mathbf{10}}}=6\lambda \hat{\lambda }
\end{equation}
This degeneracy is a consequence of integrability, and as far as the
system stays integrable should be present at higher orders of
perturbation theory. The solution for the $\mathbf{10}$ is given by
the same distribution of roots with the $u$ and $v$ roots
interchanged.

The solution of the Bethe equations for case (iii) which corresponds
to the operator in the $\mathbf{6}$, is given by
\begin{equation}\label{}
 u_1=-u_2=v_1=-v_2=\frac{1}{2}\,,\qquad
 r_1=-r_2=\pm\frac{1}{\sqrt{2}}\,\qquad
 s_1=0.
\end{equation}
Its energy is
\begin{equation}\label{}
 \gamma _{\mathbf{6}}=8\lambda \hat{\lambda }.
\end{equation}
The solutions of the Bethe equations completely agree with the
direct diagonalization of the Hamiltonian.

%%%%%%%%%%%%%%%%%%%%%%%%%%%%%%%%%%%%%%%%%%%%%%%%%%%%%%%%%%%%
\section{The Complete $R$-matrix and Integrability}

In this section we find the complete $R$-matrix for the $OSp(6|4)$ symmetry algebra.  We do this by first finding the $R$-matrix in the $SL(2|1)$ sector, a closed noncompact sector which contains fermions.  We then lift this $R$-matrix to the full $R$-matrix by showing that there there is a one to one map in the tensor product of $SL(2|1)$ and $OSp(6|4)$ representations.  We then show that that this $R$-matrix leads to the Hamiltonian in (\ref{dil}).

\subsection{The oscillator algebra and the singleton representations}

As a preliminary, we first construct the $OSp(6|4)$ generators in terms of bosonic and fermionic creation and annihilation operators.  The generators of $OSp(6|4)$ are
\begin{eqnarray}
P^{\al\beta}=P^{\beta\alpha},&&\qquad K_{\al\beta}=K_{\beta\al}\nn\\
Q^{\al I},&&\qquad {S_{\al}}^{ I}\nonumber\\
{L^{\al}}_{\beta},\ \ {L^{\g}}_{\g}=0,\qquad&& R^{IJ}=-R^{JI},\qquad D\,
\end{eqnarray}
where $\al,\beta,\g=1,2$, and $I,J=1\dots6$.  The algebra is then
\be\label{alg}
[D,P^{\al\beta}]&=&+P^{\al\beta}\qquad [D,K_{\al\beta}]=-K_{\al\beta}\nn\\
\ [D,Q^{\al I}]&=&+\half \,Q^{\al I},\qquad [D,{S_{\al}}^{ I}]=-\half\, {S^{\al}}_{ I},\nn\\
\ [P^{\al\beta},K_{\g\de}]&=&\sfrac14\left({\delta^\beta}_\delta{L^\al}_\g+{\delta^\beta}_\g{L^\al}_\delta+{\delta^\al}_\delta{L^\beta}_\g+{\delta^\al}_\g{L^\beta}_\delta\right)+\sfrac12\left({\delta^\al}_\g{\delta^\beta}_\delta+{\delta^\al}_\delta{\delta^\beta}_\g\right)D\nn\\
\{Q^{\al I},Q^{\beta J}\}&=&\,\delta^{IJ}\,P^{\al\beta},\qquad \{{S_{\al}}^{ I},{S_{\beta}}^{ J}\}=-\,\delta^{IJ}\,K_{\al\beta}\nn\\
\{Q^{\al I},{S_{\beta}}^{ J}\}&=&\half{\de^\al}_\beta R^{IJ}-\half\delta^{IJ}{L^\al}_\beta-\half{\de^\al}_\beta\,\delta^{IJ}D\nn \\
\ [ R^{IJ},Q^{\al K}]&=&\delta^{JK}Q^{\al I}-\delta^{IK}Q^{\al J},\qquad [R^{IJ},{S_\al}^K]=\delta^{JK}{S_\al}^I-\delta^{IK}{S_\al}^J,\nn\\
\ [{L^\al}_\beta,P^{\g\de}]&=&{\delta^\g}_\beta P^{\al\de}+{\delta^\de}_\beta P^{\g\al},\qquad
\ [{L^\al}_\beta,K_{\g\de}]=-{\delta^\al}_\g K_{\beta\de}-{\delta^\al}_\de K_{\g\beta}\,.
\ee
All other commutators are zero.

It is convenient to write the generators in terms of oscillators.  We introduce the bosonic oscillators $b^{\dag\al}$ and $b_\beta$ as well as the real fermionic oscillators $c^I$.  These satisfy the commutation and anticommutation relations
\be
[b_{\al},b^{\dag\beta}]={\de_\al}^\beta,\qquad \{c^I,c^J\}=\delta^{IJ}\,.
\ee
We can then write the generators as
\be\label{genosc}
P^{\al\beta}=\half\,b^{\dag\al}b^{\dag\beta},&&\qquad K_{\al\beta}=-\half\,b_{\al}b_{\beta}\,\nn\\
Q^{\al I}=\sfrac{1}{\sqrt{2}}b^{\dag\al}c^I,&&\qquad {S_{\al}}^{ I}=-\sfrac{1}{\sqrt{2}}b_\al c^I\nonumber\\
{L^{\al}}_{\beta}=b^{\dag\al}b_\beta-\half b^{\dag\g}b_\g ,\qquad&& R^{IJ}=\half(c^Ic^J-c^Jc^I),\qquad D=\half b^{\dag\g}b_\g+\half\,,
\ee
where one can easily check that this satisfies the algebra in (\ref{alg}).

We can now build two different representations of this algebra.  We first note that the 6 real fermions can be split into 3 complex fermions, $d^j=\frac{1}{\sqrt{2}}(c^{2j-1}+i\,c^{2j})$ which satisfy the anticommutation relations
\be
\{d^{j},d^{\dag k}\}=\delta^{jk}\,\qquad\{d^{\dag j},d^{\dag k}\}=\{d^{j},d^{k}\}=0\,.
\ee
Since we have 3 creation operators we can create 8 different states with the $d^{\dag j}$, half of which are fermionic.  Letting $|0\rangle$ satisfy $d^j|0\rangle=b_\al|0\rangle=0$, we have that states with an even number of $d^{\dag j}$'s acting on $|0\rangle$ are in the {\bf 4} rep of the $SO(6)$  and those with an odd number are in the $\bf\bar{4}$.  It is now clear from the form of the generators in (\ref{genosc}) that we can build two independent representations -- those with an even number of oscillators and those with an odd number.  The representation with an even (odd) number we call chiral (anti-chiral).  Both representations have  $D=\half$ for their highest weight, where for one case the highest weight is in the {\bf 4} and the other it is in the $\bf\bar4$.

We can quickly see that these representations match to the field content of the gauge theory.  For those fields in the $(N,\bar N)$ representation of the gauge group we identify
\be
D^nY^A:&&\qquad \frac{1}{2^n}\,(b^\dag)^{2n}|0\rangle, \ \ \frac{1}{2^n}\,(b^\dag)^{2n} d^{\dag j}d^{\dag k}|0\rangle,\nn\\
D^n\psi_A:&&\qquad \frac{1}{2^{n+1/2}}\,(b^\dag)^{2n+1}d^{\dag j}|0\rangle, \ \ \frac{1}{2^{n+1/2}}\,(b^\dag)^{2n+1} d^{\dag 1}d^{\dag 2}d^{\dag 3}|0\rangle,
\ee
while for those in the $(\bar N,N)$ representation the identification is
\be
D^nY^{\dag }_A:&&,\qquad \frac{1}{2^{n}}\,(b^\dag)^{2n}d^{\dag j}|0\rangle, \ \ \frac{1}{2^{n}}\,(b^\dag)^{2n} d^{\dag 1}d^{\dag 2}d^{\dag 3}|0\rangle\nn\\
D^n\psi^{\dag A}:&&\qquad \frac{1}{2^{n+1/2}}\,(b^\dag)^{2n+1}|0\rangle, \ \ \frac{1}{2^{n+1/2}}\,(b^\dag)^{2n+1} d^{\dag j}d^{\dag k}|0\rangle,
\ee
where for this latter case we change the fermion number of $|0\rangle$.  Notice that unlike the case of $\NN=4$ SYM, the field strengths $\FF$ and $\hat \FF$ don't appear as fundamental fields in the fundamental representations.  This is because there is only the Chern-Simon's kinetic term, so the field strengths are equivalent to a combination of the other fields via the equations of motion.

\subsection{The $SL(2|1)$ sector}

The smallest closed non-compact sector whose ground state is the
chiral pri\-ma\-ry operator $\,{\rm tr}\, (Y^1Y^\dagger_4)^L$ is
the $SL(2|1)$ sector\footnote{There is also a closed $SU(1,1)$
sector, where the field content has scalars on the odd sites and
fermions on the even sites, all with the same $SU(4)$ index, as well
as covariant derivatives \cite{Zwiebel:2009vb}.  We thank B. Zwiebel
for remarks on this.}.  In this case the field content is restricted
to $Y^1$ and $\psi_{4+}$ on the $(N,\bar{N})$ sites, $Y^\dag_4$ and
$\psi^{1}_{ c+}$ on the $(N,\bar{N})$ sites, as well as covariant
derivatives $D_+$.  The $+$ index is the helicity.  The states are
constructed out of one bosonic operator, say $b^\dag\equiv
b^{1\dag}$ and one fermionic oscillator $d^\dag\equiv d^{1\dag}$.

\begin{figure}[t]
\centerline{\includegraphics[height=3cm]{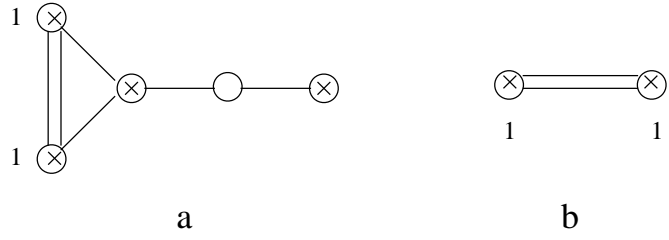}}
\caption{\label{superD}\small Super-Dynkin diagrams for (a) $OSp(6|4)$ (b) $SL(2|1)$. }
\end{figure}

A nice way to see how  $SL(2|1)$ fits into $OSp(6|4)$ is by examining their super-Dynkin diagrams.  Super-Dynkin diagrams are not unique so there is some freedom in choosing an appropriate diagram.  For $OSp(6|4)$, one such diagram has already been  shown in figure \ref{dynya}.  However, another diagram is shown in figure \ref{superD}a, where now the momentum carrying roots in the Bethe equations are fermionic.  These roots are now also coupled, with  the double line indicating that their inner product is $+2$.  The $SL(2|1)$ subgroup is taken by reducing the diagram to these two fermionic momentum carrying roots, as shown in figure \ref{superD}b.

The $SL(2|1)$ generators are
\be
J_+\equiv P^{11}=\half\,b^{\dag}b^{\dag},&&\qquad J_-\equiv K_{11}=-\half\,b\,b\,,\qquad J_0\equiv\half(D+{L^{1}}_{1})=\half b^\dag b+\sfrac14\,\nn\\
Q^{+}=\sfrac{1}{\sqrt{2}}\,b^{\dag}d^\dag,&&\qquad Q^{-}=\sfrac{1}{\sqrt{2}}\,b^{\dag}d,\qquad S^{+}=-\sfrac{1}{\sqrt{2}}\,b\,d^\dag,\qquad S^{-}=-\sfrac{1}{\sqrt{2}}\,b\,d\nonumber\\
&&H\equiv \half R^{12}=\half d^\dag d-\sfrac14\,.
\ee
$J_+$, $J_-$ and $J_0$ are the usual $SL(2)$ generators, satisfying
\be
[J_0,J_\pm]=\pm J_\pm,\qquad [J_+,J_-]=2J_0\,.
\ee
The other nontrivial commutators are
\be
[J_0,Q^\pm]=\half Q^\pm,&&\qquad[J_0,S^\pm]=-\half S^\pm,\qquad [H,Q^\pm]=\pm \half Q^\pm,\qquad [H,S^\pm]=\pm \half S^\pm\nn\\
\{Q^+,Q^-\}=J_+,&&\qquad \{S^+,S^-\}=-J_-,\qquad\{Q^+,S^-\}=H-J_0,\qquad\{Q^-,S^+\}=-H-J_0\,\nn\\
&&\qquad\qquad\qquad\, [J_+,S^{\pm}]=Q^{\pm}\,,\qquad [J_-,Q^{\pm}]=S^{\pm}\,.
\ee

The irreducible representations $\VV_{j,h}$ are labeled by the charges $(j,h)$ of the lowest weight states in the representation.  The lowest weights are annihilated by $S^\pm$ and $J_-$ and if $j>0$ then  $\VV_{j,h}$ is infinite dimensional.  If $h=- j$ ($h=+j$), then the lowest weight is also annihilated by $Q^-$ ($Q^+$).  These representations are called chiral (antichiral).  Representations that are neither chiral nor antichiral are called typical.

An important ingredient for constructing an $R$-matrix  is the tensor product of two representations.  In the case when both representations are chiral or both antichiral, the tensor product is given by
\be
\VV_{j_1,\mp j_1}\otimes\VV_{j_2,\mp j_2}=\VV_{j,\mp j}+\sum_{n=0}^\infty\VV_{j+\half+n,\mp j\pm\half}\,,
\ee
where $j=j_1+j_2$.
The first representation is chiral (anti-chiral) but the representations in the sum are typical.  If one representation is chiral and the other is antichiral, then the tensor product takes the form
\be
\VV_{j_1,\mp j_1}\otimes\VV_{j_2,\pm j_2}=\sum_{n=0}^\infty\VV_{j+n,\mp \bar{j}}\,,
\ee
where $\bar{j}=j_1-j_2$.  All representations in this sum are typical.

Let us now turn to our particular situation.  The lowest weights are the states $|0\rangle$ and $d^\dag|0\rangle$, where $b|0\rangle=d|0\rangle=0$.  Acting on these states with $J_0$ and $H$, we find
that their charges are $(\sfrac{1}{4},-\sfrac{1}{4})$ and $(\sfrac{1}{4},\sfrac{1}{4})$ respectively.  Hence $|0\rangle$ is the lowest weight of a chiral representation and $d^\dag|0\rangle$ is the lowest weight of an antichiral representation.  The tensor products are then
\be
\VV_{\quart,\mp\quart}\otimes\VV_{\quart,\mp\quart}&=&\VV_{\half,\mp\half}+\sum_{n=0}^\infty\VV_{n+1,0}\,,\nn\\
\VV_{\quart,\mp \quart}\otimes\VV_{\quart,\pm \quart}&=&\sum_{n=0}^\infty\VV_{n+\half,0}\,.
\ee

The $R$-matrix $R_{ab}(u)$ acts on the tensor product of two representations $\VV_a\otimes \VV_b$.  Since the $R$-matrix is invariant under the algebra, it can be written as a projection operator onto the representations in the tensor product,
\be
R_{ab}(u)=\sum_c R^c_{ab}(u)\PP_c
\ee
where $\PP_c$ is the projection operator onto $\VV_c$.
The universal $R$-matrix for any representation in $SL(2|1)$ was derived in \cite{Derkachov:2000ne}.  The relevant results for the $(\quart,\mp\quart)$ representations are
\be
R_{\mp\mp}^{(\half,\mp\half)}(u)&=&-\frac{\ut-\half}{\ut+\half}R_{\mp\mp}^{(1,0)}(u) \qquad R_{\mp\mp}^{(n+1,0)}(u)=(-1)^n\frac{\Gamma(\ut+n+\sfrac{3}{2})}{\Gamma(-\ut+n+\sfrac{3}{2})}\,g(\ut)\nn\\
&&R_{\mp\pm}^{(n+\half,0)}(u)=(-1)^n\frac{\Gamma(\ut+n+1)}{\Gamma(-\ut+n+1)}\,\tilde{g}(\ut)\,,
\ee
where $\ut=u/c$ with $c$ an arbitrary constant, and $g(\ut)$ and $\tilde{g}(\ut)$ are arbitrary functions.  It is convenient to choose
\be
g(\ut)=-\frac{\ut+\half}{\ut-\half}\,\frac{\Gamma(-\ut+\sfrac{3}{2})}{\Gamma(\ut+\sfrac{3}{2})},\qquad
\tilde{g}(\ut)=\frac{\Gamma(-\ut+1)}{\Gamma(\ut+1)}, \qquad c=2\,,
\ee
in which case we have,
\be
R_{\mp\mp}^{(\half,\mp\half)}(u)&=&1,\qquad R_{\mp\mp}^{(n+1,0)}(u)=\prod_{k=0}^n\frac{u+2k+1}{u-2k-1},\nn\\
R_{\mp\pm}^{(n+\half,0)}(u)&=&\prod_{k=0}^n\frac{u+2k}{u-2k}\,.
\ee

Now that we have the $R$-matrix we can construct the transfer matrices for an alternating spin-chain with a chiral representation on the odd sites and an antichiral representation on the even sites.  Following the notation in \cite{Derkachov:2000ne}, let us use $a_i$ to label the sites in the chiral representation and $\bar{a}_i$ to label sites in the anti-chiral representation.  The two distinct transfer matrices for the chain with $L$ sites are thus given by
\be
T_a(u)&=& R_{a a_1}(u)R_{a \bar{a}_1}(u)R_{a a_2}(u)R_{a \bar{a}_2}(u)\dots R_{a a_L}(u)R_{a \bar{a}_L}(u)\nn\\
T_{\ba}(u)&=& R_{\ba a_1}(u)R_{\ba \bar{a}_1}(u)R_{\ba a_2}(u)R_{a \bar{a}_2}(u)\dots R_{\ba a_L}(u)R_{a \bar{a}_L}(u)\,,
\ee
where the indices $a$ and $\ba$ refer to auxiliary spaces in the chiral and anti-chiral representations.
Defining $\tau(u)$ and $\btau(u)$ as traces over the auxiliary spaces,
\be
\tau(u)={\tr}_a T_a(u)\,,\qquad \btau(u)={\tr}_\ba T_\ba(u)\,,
\ee
the Yang-Baxter equation then guarantees the commutation relations
\be
[\tau(u),\tau(v)]=[\btau(u),\btau(v)]=[\tau(u),\btau(v)]=0,.
\ee
Hence, expanding $\tau(u)$ and $\btau(u)$ in powers of $u$ gives a commuting set of charges for the theory.

The charge we are most interested in is the Hamiltonian, $H$, which is given by
\be\label{Ham}
H=\,C\,(\tau(0)\btau(0))^{-1}\frac{d}{du}(\tau(u)\btau(u))\Big|_{u=0}\,,
\ee
where $C$ is a constant to be determined.
To explicitly construct this, we first note that
\be\label{R0}
R_{ab}(0)&=&\PP_{\half,-\half}+\sum_{j=1}^\infty (-1)^j\PP_{j,0}\nn\\
R_{\ba\bb}(0)&=&\PP_{\half,+\half}+\sum_{j=1}^\infty (-1)^j\PP_{j,0}.
\ee
In these cases, the representations $\VV_{{\tiny\half},-\half}$, $\VV_{\half,+\half}$ are symmetric representations, while $\VV_{j,0}$ is symmetric (antisymmetric) for $j$ even (odd).  Hence, we see that these operators are the exchange operators,
\be\label{R0isP}
R_{ab}(0)=P_{ab}\,,\qquad R_{\ba\bb}(0)=P_{\ba\bb}\,.
\ee
The $R$-matrix evaluated at $u=0$ between a chiral and an anti-chiral representation  is
\be\label{Rb0}
R_{a\bb}(0)&=&\sum_{j=0}^\infty (-1)^j\PP_{j+\half,0}
\ee
Using explicit indices, we write this operator as
$\MM_{I_a\bI_b}^{J_a\bJ_b}$, where $I_a$ and $\bI_b$ refers to particular elements of the chiral and anti-chiral representations.

From the results in (\ref{R0}), (\ref{R0isP}) and (\ref{Rb0}) we find
\be
\tau(0)_{I_1\bI_1\dots I_L\bI_L}^{J_1\bJ_1\dots J_L\bJ_L}&=&\MM_{I_1\bI_1}^{J_2\bJ_1}\MM_{I_2\bI_2}^{J_3\bJ_2}\dots \MM_{I_L\bI_L}^{J_1\bJ_L}\nn\\
\btau(0)_{I_1\bI_1\dots I_L\bI_L}^{J_1\bJ_1\dots J_L\bJ_L}&=&\MM_{I_2\bI_1}^{J_2\bJ_2}\MM_{I_3\bI_2}^{J_3\bJ_3}\dots \MM_{I_1\bI_L}^{J_1\bJ_1}\,.
\ee
Hence, the product of $\tau(0)$ and $\btau(0)$ is
\be
\big(\tau(0)\btau(0)\big)_{I_1\bI_1\dots I_L\bI_L}^{J_1\bJ_1\dots J_L\bJ_L}
&=&(\MM^2)_{I_1\bI_1}^{J_2\bJ_2}(\MM^2)_{I_2\bI_2}^{J_3\bJ_3}\dots (\MM^2)_{I_L\bI_L}^{J_1\bJ_1}\nn\\
&=&\delta_{I_1}^{J_2} \delta_{\bI_1}^{\bJ_2}\delta_{I_2}^{J_3} \delta_{\bI_2}^{\bJ_3}\dots
\delta_{I_L}^{J_1} \delta_{\bI_L}^{\bJ_1}\,,
\ee
where we used (\ref{Rb0}) to get to the second line.  Therefore, this operator shifts every index over by two sites.

The first derivatives of the $R$-matrices we write in terms of two operators $\AA$ and $\BB$
\be
\AA&\equiv& \,R_{ab}'(0)=\,R_{\ba\bb}'(0)=\sum_{j=1}^\infty(-1)^j\left(\sum_{k=1}^j\frac{2}{2k-1}\right)\PP_{j,0}\nn\\
&=&
\sum_{j=1}^\infty(-1)^j\big(2\,h(2j-1)-h(j-1)\big)\PP_{j,0}\nn\\
 \BB&\equiv&\,R_{ab}'(0)=\sum_{j=1}^\infty (-1)^j h(j)\,\PP_{j+\half,0}\,,
\ee
where $h(j)$ is the harmonic sum
\be
h(j)\equiv \sum_{k=1}^j \frac{1}{k}\,.
\ee
The Hamiltonian is then found to be
\be\label{ham_r}
H&=&C\sum_{\ell=1}^L\left(\MM_{I_\ell\bI_\ell}^{K\bK}\AA^{J_{\ell+1}L}_{KI_{\ell+1}}\MM_{\bK L}^{\bJ_\ell J_\ell}+\BB_{I_\ell\bI_\ell}^{K\bK}\MM_{K\bK}^{J_\ell\bJ_\ell}\right)\nn\\
&&+\qquad C\sum_{\ell=1}^L\left(\MM_{\bI_\ell I_{\ell+1}}^{\bK K}\AA^{\bJ_{\ell+1}\bL}_{\bK\bI_{\ell+1}}\MM_{K\bL}^{ J_\ell\bJ_\ell}+\BB_{I_{\ell+1}\bI_\ell}^{K\bK}\MM_{K\bK}^{J_{\ell+1}\bJ_\ell}\right)
\ee
It's structure has next to nearest neighbor form.

\subsection{The lift to $OSp(6|4)$}

We can construct all unitary representations of $OSp(6|4)$ using bosonic and fermionic oscillators \cite{Gunaydin:1988kz}.  This is accomplished by writing the generators in Jordan form.  In particular, the algebra can be decomposed into the vector space $L_{-1}\oplus L_0\oplus L_{+1}$, with the generators in each of these subspaces labeled by
\be
S_{AB}\in L_{-1}\,,\qquad S^{AB}\in L_{+1}\,,\qquad M^A_B\in L_0\,.
\ee
The indices $A$ and $B$ run from $1$ to $5$, with $A=\alpha$, $\al=1,2$ for the bosonic indices and $A=2+i$,  $i=1,2,3$ for the fermionic indices.  The elements in $L_0$ make up the compact $U(2|3)$ subalgebra.

We can then construct sets of oscillators
\be
C_{A,r}=(b_r^1,b_r^2,d_r^1,d_r^2,d_r^3)\,,\qquad C_r^A=(b_r^{1\dag},b_r^{1\dag},d_r^{1\dag},d_r^{1\dag},d_r^{1\dag})\,.
\ee
  For our purposes where we consider the tensor product of two singleton representations, we let $r=1,2$.  If we then define
\be\label{opscomb}
\theta_A=\frac{1}{\sqrt{2}}\left(C_{A,1}+i\,C_{A,2}\right)&&\qquad\theta^A=\frac{1}{\sqrt{2}}\left(C^A_1-i\,C^A_2\right)\nn\\
\chi^A=\frac{1}{\sqrt{2}}\left(C_{A,1}-i\,C_{A,2}\right)&&\qquad\chi^A=\frac{1}{\sqrt{2}}\left(C^A_1+i\,C^A_2\right)\,,
\ee
we can then write the elements of the algebra as
\be
S_{AB}=\theta_A\chi_B+\chi_A\theta_B\qquad S^{AB}=\theta^A\chi^B+\chi^A\theta^B\nn\\
M^A_B=\theta^A\theta_B+(-1)^{\deg(A)\deg(B)}\chi_B\chi^A\,,
\ee
where $\deg(A)$ is $0$ ($1$) for bosonic (fermionic) indices.

The irreducible representations are labeled by the lowest weights, that is those states that are annihilated by the elements of $L_{-1}$.  These states themselves are representations of the $U(2|3)$ subalgebra, hence an irreducible representation of $OSp(6|4)$ is given by the corresponding irreducible representation of $U(2|3)$.  It is not hard to show that the lowest weights have the form
\be\label{lowwgt}
|0\rangle,\qquad && \left(\theta^{A}\chi^{B}-\chi^{A}\theta^{B}\right)|0\rangle\nn\\
\theta^A|0\rangle,\qquad \theta^{A_1}\theta^{A_2}|0\rangle,&&\qquad\dots\qquad
\theta^{A_1}\dots\theta^{A_k}|0\rangle,\qquad\dots\nn\\
\chi^A|0\rangle,\qquad \chi^{A_1}\chi^{A_2}|0\rangle,&&\qquad\dots\qquad
\chi^{A_1}\dots\chi^{A_k}|0\rangle,\qquad\dots
\ee
The corresponding $U(2|3)$ representations are given by the singlet {\bf 1} and the graded antisymmetric representation, $\tiny\young(/,/)$ for the states in the top line of (\ref{lowwgt}), while the representations in the second line are the graded symmetric product of $k$ elements $\tiny\young(/)\young(/)\dots\young(/)\young(/)$\,.  The representations in the third line are isomorphic to those in the second line, so we may  write an irreducible representation as   linear combinations of states in the second and third line.  In particular, we choose the two combinations $\theta^{A_1}\dots\theta^{A_k}\pm\chi^{A_1}\dots\chi^{A_k}$ for our irreducible representations.  Then the tensor product of two chiral representations gives
\be
\mbox{${\bf 1}\,+\,{\small \young(/)\young(/)}{\,}_+ \,+\,{\small\young(/)\young(/)\young(/)\young(/)}{\,}_+\,+\,\dots$}\,,
\ee
while the tensor product of two anti-chiral representations is
\be
\mbox{${\small\young(/,/)}\,\,+\,\,{\small \young(/)\young(/)}{\,}_- \,+\,{\small\young(/)\young(/)\young(/)\young(/)}{\,}_-\,+\,\dots$}\,.
\ee
The subscript on the symmetric representations refers to which combination of $\theta^A$ and $\chi^A$ we choose.  Note however, that these are the same representations.
The tensor product of a chiral and an antichiral representation is
\be
\mbox{${\small\young(/)}{\,}_+ \,+\,{\small\young(/)\young(/)\young(/)}{\,}_+\,+\,\dots$}\,,
\ee
while the product of the anti-chiral and the chiral representation reverses the signs in the subscripts.

Under the exchange $1\leftrightarrow 2$, we have that $\theta^A\to -i\chi^A$, $\chi^A\to +i\theta^A$.  Thus, for even $k$
\be
\big(\theta^{A_1}\dots\theta^{A_k}\pm\chi^{A_1}\dots\chi^{A_k}\big)\to \pm(-1)^{k/2}\big(\theta^{A_1}\dots\theta^{A_k}\pm\chi^{A_1}\dots\chi^{A_k}\big)\,.
\ee
Choosing $|0\rangle\to +|0\rangle$ under the exchange in the chiral-chiral tensor product and {$|0\rangle\to -|0\rangle$} in the antichiral-antichiral tensor product, we see that the representations in these tensor products are exchange eigenstates, with eigenvalue $(-1)^{k/2}$.

Let us now consider the $SL(2|1)\simeq OSp(2|2)$ subgroup discussed in the last section.  In this case the representations can be labeled by the representations of the $U(1|1)$ compact subgroup.  However, the super Young tableaux have the same form as in the $U(2|3)$, hence there is a one to one map between the tensor products of  chiral or antichiral representations in $SL(2|1)$ and in $OSp(6|4)$.  Note that {\bf 1} is the chiral representation and $\tiny\young(/,/)$ is the anti-chiral representation.  The representation with $k$ graded symmetric boxes corresponds to the  representation with charges $j=k/2, h=0$.  Notice further that this matches the symmetries under the exchange.  Hence, the $R$-matrix for $OSp(6|4)$ has precisely the same form as in the previous section, with the $SL(2|1)$ projections  replaced with the corresponding $OSp(6|4)$ projections.

\subsection{Subsectors}

\subsubsection{$SU(4)$}

For this subsector we only have the symmetric ({\bf 10} or ${\bf\overline{10}}$) and anti-symmetric ({\bf 6}) representations in the tensor product of two chiral or two anti-chiral representations, which corresponds to the $j=1/2$ and $j=1$ representations  respectively.  Hence, in this case the $R$-matrix is
\be\label{s06rm1}
R_{ab}&=&P_{10}+\frac{u+1}{u-1}\,P_{6}\nn\\
&=&\half(1+P_{ab})+\frac{u+1}{u-1}\,\half(1-P_{ab})\,=\,\frac{1}{u-1}(u-P_{ab})\,.
\ee
Similarly,
\be\label{s06rm2}
 R_{\ba\bb}=P_{\,\overline{10}}+\frac{u+1}{u-1}\,P_{{6}}=\,\frac{1}{u-1}(u-P_{\ba\bb}).
 \ee
For the chiral-antichiral case we have only projections onto the adjoint ({\bf 15}), which corresponds to $j=1/2$, and the singlet ({\bf 1}), which is $j=3/2$.  Hence, the $R$-matrix is
\be\label{s06rm3}
R_{a\bb}&=&P_{15}+\frac{u+2}{u-2}\,P_1=(1-\quarter K_{a\bb})+\frac{u+2}{u-2}\,\quarter K_{a\bb}\nn\\
&=&\frac{1}{u-2}\,(u+K_{a\bb}-2)\,.
\ee
These are the $R$-matrices previously given in \cite{Minahan:2008hf}, with  the $R_{a\bb}$ matrix shifted so that it satisfies a standard Yang-Baxter equation.  There is also an overall function in front of each $R$-matrix, but this does not affect the Yang-Baxter equation and only shifts the energy by a constant amount.
In fact, with this finite shift the Hamiltonian acting on a chiral primary gives zero.

With these $R$-matrices, we find that
\be
\MM=1-K/2\,,\qquad \AA= -(1-P)\,,\qquad \BB=-\quart K\,.
\ee
A quick calculation  gives the Hamiltonian in (\ref{dilopbos}) and fixes the coefficient in (\ref{Ham}) to $C=\la\,\hat\la$.

\subsubsection{Scalars and one fermion}

The tensor product of a fermion $\psi_A$ and a scalar $Y^B$ decomposes into an $SU(4)$ {\bf 15} or {\bf 1}.   Symmetrizing or antisymmetrizing over their positions, we find that the {\bf 15${}_s$} is in the $OSp(6|4)$ representation labeled by {\bf 1}, {\bf 15${}_a$} and {\bf 1${}_a$} are in $\small\young(/)\young(/)$\,, and {\bf 1${}_s$} is in $\small\young(/)\young(/)\young(/)\young(/)$\,.  If we define the action of the exchange $\hat{P}$ and trace operators to be
\be
\hat P\,\psi_A \cdots Y^B = Y^B\cdots \psi_A\,,\qquad K\, \psi_A \cdots Y^B=\delta_A^B\, \psi_D\cdots Y^D\,,
\ee
Then the action of the $R$-matrix on $\psi_A\cdots Y^B$ is
\be
&&R(u)\,\psi_A\cdots Y^B\nn\\
&&=\left[\half(1+\hat P)(1-\quart K)+\frac{u+1}{u-1}\half(1-\hat P)+\frac{(u+1)(u+3)}{(u-1)(u-3)}\half(1+\hat P)\quart K\right]\psi_A \cdots Y^B\,.\nn\\
\ee
Thus
\be\label{1fr1}
R(0)&=&\hat P\,,\nn\\
\AA=R'(0)\,&=&-(1-\hat P)+\sfrac{1}{3}(1+\hat P)K\,.
\ee

Likewise, the product of $\psi_AY^\dag_B$ can be expressed as
\be
\psi_AY^\dag_B=\psi_{\{A,}Y^\dag_{B\}}&+&\half\left(\psi_{[A,}Y^\dag_{B]}+\half\varepsilon_{ABCD}Y^C\psi^{D}_c\right)\nn\\
&+&\half\left(\psi_{[A,}Y^\dag_{B]}-\half\varepsilon_{ABCD}Y^C\psi^{D}_c\right)\,.
\ee
The first line of the decomposition is in $\small\young(/)$\,, while the second line is in $\small\young(/)\young(/)\young(/)$\,.  The decomposition is the same if the $SU(4)$ indices are raised.
If we define the action of $\hat K$ as
\be
\hat K\, \psi_A Y^\dag_B=\half\varepsilon_{ABCD}\, Y^C\psi^{D}_c\,,
\ee
then we can write the decomposition as
\be
1=\half(1+P)+\quart(1+\hat K)(1-P)+\quart(1-\hat K)(1-P)\,,
\ee
where $P$ exchanges $SU(4)$ indices.  Hence, the corresponding $R$-matrix is
\be
R(u)=\sfrac{3}{4}+\sfrac{1}{4}P+\sfrac{1}{4}\hat K(1-P)+\frac{u+2}{u-2}(1-\hat K)(1-P)\,,
\ee
which gives
\be\label{1fr2}
\MM=R(0)&=&\half(1+P)+\half\hat K(1-P)\nn\\
\BB=R'(0)&=&-\quart(1-\hat K)(1-P)\,.
\ee

Using the results in (\ref{1fr1}) and (\ref{1fr2}), as well as those in (\ref{s06rm1}-\ref{s06rm3}), we find that
\be\label{HpsiYY}
\MM\AA\MM\circ \psi_A Y^\dag_B Y^C=&&
\frac{3}{4}\,\psi_AY^\dag_BY^C+\frac{1}{4}\, \psi _BY^\dagger _A Y^C-\frac{1}{2}\,Y^CY^\dagger _B\psi _A-\frac{1}{2}\,Y^CY^\dagger _A\psi _B\nn\\
&& +\delta _B^C\left(\frac{1}{6}\, Y^DY^\dagger _A\psi _D
 +\frac{1}{6}\,Y^DY^\dagger _D\psi _A+\frac13\,\psi _AY^\dagger _DY^D-\frac{1}{6}\,\psi _DY^\dagger _AY^D\right)\nn\\
&&+\delta _A^C \left(\frac{1}{6}\,Y^DY^\dagger _D\psi _B +\frac{1}{6}\,Y^DY^\dagger _B\psi _D
-\frac{1}{6}\,\psi _BY^\dagger _DY^D+\frac{1}{3}\,\psi _DY^\dagger _BY^D\right)\nn\\
&&+\left(\frac{1}{6}\,\delta _B^C\,\varepsilon_{ADEF}-\frac{1}{3}\,\delta _A^C\,\varepsilon_{BDEF}\right)Y^D{\psi }_c^EY^F\nn\\
&&+\varepsilon_{ABDE}\left(\frac{1}{2}\,Y^D\psi^C_cY^E-\frac{1}{4}\,Y^D\psi^E_cY^C\right)
 \ee
where we used  the  identity
\be\label{identity}
\varepsilon_{ABDE}(Y^D\psi^C_cY^E-Y^C\psi^D_cY^E)+\left(\delta^C_B\,\varepsilon_{ADEF}-\delta^C_A\,\varepsilon_{BDEF}\right)Y^D\psi^E_cY^F=\varepsilon_{ABDE}Y^D\psi^E_cY^C\,.\ \ \ \ee
We also have
\be\label{HYYpsi}
\MM\AA\MM\circ Y^A Y^\dag_B \psi_C=&&
 Y^A Y^\dag_B \psi_C-\frac{1}{2}\,\psi _CY^\dagger _BY^A
 -\frac{1}{2}\,\psi _BY^\dagger _CY^A
 \nonumber \\ &&
 +\delta _B^A\left(\frac{1}{6}\,\psi _DY^\dagger _CY^D+\frac{1}{6}\,\psi _CY^\dagger _DY^D
 +\frac{1}{12}\,Y^DY^\dagger _D\psi _C-\frac{1}{6}\,Y^DY^\dagger _C\psi _D\right)
  \nonumber \\ &&
 +\delta _C^A\left(\frac{1}{6}\,\psi _BY^\dagger _DY^D +\frac{1}{6}\,\psi _DY^\dagger _BY^D
 -\frac{1}{6}\,Y^DY^\dagger _D\psi _B+\frac{1}{3}Y^DY^\dagger _B\psi _D\right)
 \nonumber \\ &&
 +\left(-\frac{1}{6}\,\delta _B^A\,\varepsilon _{CDEF} +\frac{1}{3}\delta _C^A\,\varepsilon _{BDEF}\right)Y^D{\psi }_c^EY^F
 \nonumber \\ &&
 +\varepsilon _{BCDE}\left(-\frac{1}{2}\,Y^A{\psi }_c^DY^E
 +\frac{1}{2}\,Y^D{\psi }_c^AY^E\right)
\end{eqnarray}
where we used the transpose of (\ref{identity}), and
\begin{eqnarray}\label{HYpsiY}
\MM\AA\MM \circ Y^\dagger _A\psi _BY^\dagger _C&=&\frac{13}{12}\,Y^\dagger _A\psi _BY^\dagger _C-\frac{1}{3}\,Y^\dagger _A\psi _CY^\dagger _B -\frac{1}{12}\,Y^\dagger _B\psi _AY^\dagger _C
\nn\\ &&
-\frac{1}{6}\,Y^\dagger _B\psi _CY^\dagger _A
 -\frac{1}{6}\,Y^\dagger _C\psi _AY^\dagger _B -\frac{1}{3}Y^\dagger _C\psi _BY^\dagger _A \nonumber \\ &&
 -\frac{1}{2}\,\varepsilon _{BCDE}Y^\dagger _AY^D{\psi}_c^E
 +\varepsilon _{ACDE}\left(\frac{1}{2}\,Y^\dagger _BY^D{\psi }_c^E +\frac{1}{4}\,{\psi }_c^DY^EY^\dagger _B\right)
 \nn\\&&
 -\frac{1}{4}\,\varepsilon _{ABDE}\,{\psi }_c^DY^EY^\dagger _C
 \nonumber \\ &&
 +\varepsilon _{ABCD}\left(\frac{1}{6}\,Y^\dagger _EY^E{\psi }_c^D-\frac{1}{6}\,{\psi }_c^DY^EY^\dagger _E -\frac{1}{3}\,Y^\dagger _EY^D{\psi}_c^E+\frac{1}{3}\,{\psi }_c^EY^DY^\dagger _E\right)\nn\\
\end{eqnarray}
where we used
\be
\varepsilon _{ABCD}\left(Y^\dagger _EY^D{\psi }_c^E-Y^\dagger _EY^E{\psi }_c^D\right)+
\left(\varepsilon_{ABDE}\,Y^\dagger_C +\varepsilon_{BCDE}\,Y^\dagger_A+\varepsilon_{CADE}\,Y^\dagger_A\right)Y^D{\psi }_c^E=0\,\nn
\ee
and its transpose.

We also have
\be\label{HpsiY}
 \BB\MM\circ\psi _AY^\dagger _B&=&\frac{1}{4}\,\psi _AY^\dagger _B
-\frac{1}{4} \psi _BY^\dagger _A
 -\frac{1}{4}\,\varepsilon _{ABCD}\,Y^C{\psi}_c
 ^D \\
 \label{HYpsi}
 \BB\MM\circ Y^\dagger _A\psi _B&=&\frac{1}{4}\,Y^\dagger _A\psi _B-
\frac{1}{4}\, Y^\dagger _B\psi _A
 -\frac{1}{4}\,\varepsilon _{ABCD}\,{\psi}_c^CY^D\\
 \label{HYY}
  \BB\MM\circ Y^AY^\dagger _B&=&\frac{1}{4}\,\delta^A_B\,Y^DY^\dagger _D\,,
\end{eqnarray}
 where it is necessary to include (\ref{HYY}) because it is not paired with a term having the form $\MM\AA\MM\circ Y^AY^\dagger _B Y^C$, as would be the case for a strictly bosonic chain.

Summing over (\ref{HpsiYY}) and (\ref{HYYpsi})-(\ref{HYY}), one readily sees that this gives $\Gamma_{\rm ferm}$ in (\ref{dil}) with the same diagonal piece.

\section{Discussion}

In  this paper we constructed the two-loop Hamiltonian for the full
$OSp(6|4)$ group in terms of  projectors onto irreducible
representations.  The Hamiltonian has next to nearest neighbor form
and the results agree with explicit two loop calculations for
fermionic operators.

At present, we are still  puzzled by the apparent unbroken parity
symmetry of the ABJ model, at least in the planar limit. The results
of Zwiebel \cite{Zwiebel:2009vb} seem to suggest that the two-loop
equivalence of the ABJ and ABJM models is a consequence of
supersymmetry, since in \cite{Zwiebel:2009vb} the two-loop
Hamiltonian is constructed algebraically by imposing the
supersymmetry constraints on the most general structure consistent
with planarity of the Feynman diagrams. The parity comes out
automatically in Zwiebel's construction \cite{Zwiebel:2009vb}.
It may happen that the ABJM and ABJ models are equivalent at the
planar level, up to replacement of $\lambda ^2$ in ABJM by $\lambda
\hat{\lambda }$ in ABJ, even at higher loop orders. But
it seems just as  likely that parity is an accidental symmetry of the two-loop
approximation and is broken at higher loops if $\lambda \neq\hat{\lambda }$.
It remains to be seen if the spin chain stays
integrable for generic $\lambda $ and $\hat{\lambda }$.

%%%%%%%%%%%%%%%%%%%%%%%%%%%%%%%%
\subsection*{Acknowledgments}
We would like to thank V.~Kazakov and M.~Staudacher for interesting
discussions. We also thank B. Zwiebel for comments on the manuscript.
The research of J.~A.~M. is supported in part by the
Swedish research council and the STINT foundation. The research of
W.~S. is supported in part by the ANR (CNRS-USAR) contract
05-BLAN-0079-01.  The work of K.~Z. was supported in part by the
Swedish Research Council under the contract 621-2007-4177, in part
by the RFFI grant  06-02-17383, and in part by the the grant for
support of scientific schools NSH-3036.2008.2. J.~A.~M. thanks the
CTP at MIT and ENS for kind hospitality during the course of this
work.
%%%%%%%%%%%%%%%%%%%%%%%%%%%%%%%%%%

%\appendix

%%%%%%%%%%%%%%%%%%%%%%%%%%%%%%%%%%%%%%%%%%%%%%%%%%%%%%%%%%%%%%%%%%%%%%%%%%%%%%%%

\bibliographystyle{nb}
\bibliography{refs}

\begin{thebibliography}{10}
\ifx\href\asklfhas\newcommand{\href}[2]{#2}\fi
\raggedright
\small
\parskip 0pt

%%CITATION = 0806.1218;%%
\bibitem{Aharony:2008ug}
O.~Aharony, O.~Bergman, D.~L.~Jafferis and J.~Maldacena,
\textit{``{N=6 superconformal Chern-Simons-matter theories, M2-branes and their
  gravity duals}''},
\href{http://arXiv.org/abs/0806.1218}{\texttt{0806.1218}}.
%
%%CITATION = 0806.1519;%%
\bibitem{Benna:2008zy}
M.~Benna, I.~Klebanov, T.~Klose and M.~Smedback,
\textit{``{Superconformal Chern-Simons Theories and $AdS_4/CFT_3$
  Correspondence}''},
\textsf{JHEP~0809,~072~(2008)},
\href{http://arXiv.org/abs/0806.1519}{\texttt{0806.1519}}.
%
%%CITATION = 0806.4977;%%
\bibitem{Hosomichi:2008jb}
K.~Hosomichi, K.-M.~Lee, S.~Lee, S.~Lee and J.~Park,
\textit{``{N=5,6 Superconformal Chern-Simons Theories and M2-branes on
  Orbifolds}''},
\textsf{JHEP~0809,~002~(2008)},
\href{http://arXiv.org/abs/0806.4977}{\texttt{0806.4977}}.
%
%%CITATION = HEP-TH/9209005;%%
\bibitem{Chen:1992ee}
W.~Chen, G.~W.~Semenoff and Y.-S.~Wu,
\textit{``{Two loop analysis of nonAbelian Chern-Simons theory}''},
\textsf{Phys.~Rev.~D46,~5521~(1992)},
\href{http://arXiv.org/abs/hep-th/9209005}{\texttt{hep-th/9209005}}.
%
%%CITATION = 0704.3740;%%
\bibitem{Gaiotto:2007qi}
D.~Gaiotto and X.~Yin,
\textit{``{Notes on superconformal Chern-Simons-matter theories}''},
\textsf{JHEP~0708,~056~(2007)},
\href{http://arXiv.org/abs/0704.3740}{\texttt{0704.3740}}.
%
%%CITATION = 0806.3951;%%
\bibitem{Minahan:2008hf}
J.~A.~Minahan and K.~Zarembo,
\textit{``{The Bethe ansatz for superconformal Chern-Simons}''},
\textsf{JHEP~0809,~040~(2008)},
\href{http://arXiv.org/abs/0806.3951}{\texttt{0806.3951}}.
%
%%CITATION = 0807.2063;%%
\bibitem{Bak:2008cp}
D.~Bak and S.-J.~Rey,
\textit{``{Integrable Spin Chain in Superconformal Chern-Simons Theory}''},
\textsf{JHEP~0810,~053~(2008)},
\href{http://arXiv.org/abs/0807.2063}{\texttt{0807.2063}}.
%
%%CITATION = 0806.4589;%%
\bibitem{Gaiotto:2008cg}
D.~Gaiotto, S.~Giombi and X.~Yin,
\textit{``{Spin Chains in N=6 Superconformal Chern-Simons-Matter Theory}''},
\href{http://arXiv.org/abs/0806.4589}{\texttt{0806.4589}}.
%
%%CITATION = 0807.0777;%%
\bibitem{Gromov:2008qe}
N.~Gromov and P.~Vieira,
\textit{``{The all loop AdS4/CFT3 Bethe ansatz}''},
\href{http://arXiv.org/abs/0807.0777}{\texttt{0807.0777}}.
%
%%CITATION = 0807.1924;%%
\bibitem{Ahn:2008aa}
C.~Ahn and R.~I.~Nepomechie,
\textit{``{N=6 super Chern-Simons theory S-matrix and all-loop Bethe ansatz
  equations}''},
\textsf{JHEP~0809,~010~(2008)},
\href{http://arXiv.org/abs/0807.1924}{\texttt{0807.1924}}.
%
%%CITATION = 0806.4940;%%
\bibitem{Arutyunov:2008if}
G.~Arutyunov and S.~Frolov,
\textit{``{Superstrings on $AdS_4 \times CP^3$ as a Coset Sigma-model}''},
\textsf{JHEP~0809,~129~(2008)},
\href{http://arXiv.org/abs/0806.4940}{\texttt{0806.4940}}.
%
%%CITATION = 0806.4948;%%
\bibitem{Stefanski:2008ik}
j.~Stefanski,~B.,
\textit{``{Green-Schwarz action for Type IIA strings on $AdS_4\times CP^3$}''},
\textsf{Nucl.~Phys.~B808,~80~(2009)},
\href{http://arXiv.org/abs/0806.4948}{\texttt{0806.4948}}.
%
%%CITATION = 0807.0437;%%
\bibitem{Gromov:2008bz}
N.~Gromov and P.~Vieira,
\textit{``{The AdS4/CFT3 algebraic curve}''},
\href{http://arXiv.org/abs/0807.0437}{\texttt{0807.0437}}.
%
%%CITATION = 0806.3391;%%
\bibitem{Nishioka:2008gz}
T.~Nishioka and T.~Takayanagi,
\textit{``{On Type IIA Penrose Limit and N=6 Chern-Simons Theories}''},
\textsf{JHEP~0808,~001~(2008)},
\href{http://arXiv.org/abs/0806.3391}{\texttt{0806.3391}}.
%
%%CITATION = 0806.4959;%%
\bibitem{Grignani:2008is}
G.~Grignani, T.~Harmark and M.~Orselli,
\textit{``{The SU(2) x SU(2) sector in the string dual of N=6 superconformal
  Chern-Simons theory}''},
\href{http://arXiv.org/abs/0806.4959}{\texttt{0806.4959}}.
%
%%CITATION = 0807.0205;%%
\bibitem{Grignani:2008te}
G.~Grignani, T.~Harmark, M.~Orselli and G.~W.~Semenoff,
\textit{``{Finite size Giant Magnons in the string dual of N=6 superconformal
  Chern-Simons theory}''},
\textsf{JHEP~0812,~008~(2008)},
\href{http://arXiv.org/abs/0807.0205}{\texttt{0807.0205}}.
%
%%CITATION = 0807.1527;%%
\bibitem{Astolfi:2008ji}
D.~Astolfi, V.~G.~M.~Puletti, G.~Grignani, T.~Harmark and M.~Orselli,
\textit{``{Finite-size corrections in the SU(2) x SU(2) sector of type IIA
  string theory on $AdS_4 x CP^3$}''},
\href{http://arXiv.org/abs/0807.1527}{\texttt{0807.1527}}.
%
%%CITATION = 0807.0802;%%
\bibitem{Chen:2008qq}
B.~Chen and J.-B.~Wu,
\textit{``{Semi-classical strings in $AdS_4*CP^3$}''},
\textsf{JHEP~0809,~096~(2008)},
\href{http://arXiv.org/abs/0807.0802}{\texttt{0807.0802}}.
%
%%CITATION = 0807.2559;%%
\bibitem{Lee:2008ui}
B.-H.~Lee, K.~L.~Panigrahi and C.~Park,
\textit{``{Spiky Strings on $AdS_4 \times {\bf CP}^3$}''},
\textsf{JHEP~0811,~066~(2008)},
\href{http://arXiv.org/abs/0807.2559}{\texttt{0807.2559}}.
%
%%CITATION = 0807.2861;%%
\bibitem{Shenderovich:2008bs}
I.~Shenderovich,
\textit{``{Giant magnons in $AdS_4/CFT_3$: dispersion, quantization and
  finite-size corrections}''},
\href{http://arXiv.org/abs/0807.2861}{\texttt{0807.2861}}.
%
%%CITATION = 0807.3134;%%
\bibitem{Ahn:2008hj}
C.~Ahn, P.~Bozhilov and R.~C.~Rashkov,
\textit{``{Neumann-Rosochatius integrable system for strings on $AdS_4 x
  CP^3$}''},
\textsf{JHEP~0809,~017~(2008)},
\href{http://arXiv.org/abs/0807.3134}{\texttt{0807.3134}}.
%
%%CITATION = 0808.3057;%%
\bibitem{Rashkov:2008rm}
R.~C.~Rashkov,
\textit{``{A note on the reduction of the AdS4 x CP3 string sigma model}''},
\textsf{Phys.~Rev.~D78,~106012~(2008)},
\href{http://arXiv.org/abs/0808.3057}{\texttt{0808.3057}}.
%
%%CITATION = 0809.5106;%%
\bibitem{Ryang:2008rc}
S.~Ryang,
\textit{``{Giant Magnon and Spike Solutions with Two Spins in AdS4xCP3}''},
\textsf{JHEP~0811,~084~(2008)},
\href{http://arXiv.org/abs/0809.5106}{\texttt{0809.5106}}.
%
%%CITATION = 0810.0704;%%
\bibitem{Bombardelli:2008qd}
D.~Bombardelli and D.~Fioravanti,
\textit{``{Finite-Size Corrections of the $\mathbb{CP}^3$ Giant Magnons: the
  L\'{u}scher terms}''},
\href{http://arXiv.org/abs/0810.0704}{\texttt{0810.0704}}.
%
%%CITATION = 0810.1246;%%
\bibitem{Lukowski:2008eq}
T.~Lukowski and O.~O.~Sax,
\textit{``{Finite size giant magnons in the SU(2) x SU(2) sector of $AdS_4 x
  CP^3$}''},
\textsf{JHEP~0812,~073~(2008)},
\href{http://arXiv.org/abs/0810.1246}{\texttt{0810.1246}}.
%
%%CITATION = 0810.1915;%%
\bibitem{Ahn:2008tv}
C.~Ahn and R.~I.~Nepomechie,
\textit{``{An alternative S-matrix for N=6 Chern-Simons theory ?}''},
\href{http://arXiv.org/abs/0810.1915}{\texttt{0810.1915}}.
%
%%CITATION = 0810.2079;%%
\bibitem{Ahn:2008wd}
C.~Ahn and P.~Bozhilov,
\textit{``{Finite-size Effect of the Dyonic Giant Magnons in N=6 super
  Chern-Simons Theory}''},
\href{http://arXiv.org/abs/0810.2079}{\texttt{0810.2079}}.
%
%%CITATION = 0810.3516;%%
\bibitem{Jain:2008mt}
S.~Jain and K.~L.~Panigrahi,
\textit{``{Spiky Strings in AdS$_4 \times$ {\bf CP}$^3$ with Neveu- Schwarz
  Flux}''},
\textsf{JHEP~0812,~064~(2008)},
\href{http://arXiv.org/abs/0810.3516}{\texttt{0810.3516}}.
%
%%CITATION = 0811.2150;%%
\bibitem{Kristjansen:2008ib}
C.~Kristjansen, M.~Orselli and K.~Zoubos,
\textit{``{Non-planar ABJM Theory and Integrability}''},
\href{http://arXiv.org/abs/0811.2150}{\texttt{0811.2150}}.
%
%%CITATION = 0811.2423;%%
\bibitem{Abbott:2008qd}
M.~C.~Abbott and I.~Aniceto,
\textit{``{Giant Magnons in AdS4 x CP3: Embeddings, Charges and a
  Hamiltonian}''},
\href{http://arXiv.org/abs/0811.2423}{\texttt{0811.2423}}.
%
%%CITATION = 0811.2775;%%
\bibitem{Sundin:2008vt}
P.~Sundin,
\textit{``{The AdS(4) x CP(3) string and its Bethe equations in the near plane
  wave limit}''},
\href{http://arXiv.org/abs/0811.2775}{\texttt{0811.2775}}.
%
%%CITATION = 0812.2645;%%
\bibitem{Bak:2008xq}
D.~Bak,
\textit{``{Zero Modes for the Boundary Giant Magnons}''},
\href{http://arXiv.org/abs/0812.2645}{\texttt{0812.2645}}.
%
%%CITATION = 0812.2727;%%
\bibitem{Lee:2008yq}
B.-H.~Lee and C.~Park,
\textit{``{Unbounded Multi Magnon and Spike}''},
\href{http://arXiv.org/abs/0812.2727}{\texttt{0812.2727}}.
%
%%CITATION = 0807.4924;%%
\bibitem{Aharony:2008gk}
O.~Aharony, O.~Bergman and D.~L.~Jafferis,
\textit{``{Fractional M2-branes}''},
\textsf{JHEP~0811,~043~(2008)},
\href{http://arXiv.org/abs/0807.4924}{\texttt{0807.4924}}.
%
%%CITATION = 0808.0170;%%
\bibitem{Bak:2008vd}
D.~Bak, D.~Gang and S.-J.~Rey,
\textit{``{Integrable Spin Chain of Superconformal U(M)xU(N) Chern- Simons
  Theory}''},
\textsf{JHEP~0810,~038~(2008)},
\href{http://arXiv.org/abs/0808.0170}{\texttt{0808.0170}}.
%
%%CITATION = 0811.1566;%%
\bibitem{Gomis:2008jt}
J.~Gomis, D.~Sorokin and L.~Wulff,
\textit{``{The complete $AdS_4\times CP^3$ superspace for the type IIA
  superstring and D-branes}''},
\href{http://arXiv.org/abs/0811.1566}{\texttt{0811.1566}}.
%
%%CITATION = NUPHA,B379,602;%%
\bibitem{Zamolodchikov:1992zr}
A.~B.~Zamolodchikov and A.~B.~Zamolodchikov,
\textit{``{Massless factorized scattering and sigma models with topological
  terms}''},
\textsf{Nucl.~Phys.~B379,~602~(1992)}.
%
%%CITATION = 0807.3965;%%
\bibitem{McLoughlin:2008ms}
T.~McLoughlin and R.~Roiban,
\textit{``{Spinning strings at one-loop in $AdS_4 \times P^3$}''},
\href{http://arXiv.org/abs/0807.3965}{\texttt{0807.3965}}.
%
%%CITATION = 0807.4400;%%
\bibitem{Alday:2008ut}
L.~F.~Alday, G.~Arutyunov and D.~Bykov,
\textit{``{Semiclassical Quantization of Spinning Strings in $AdS_4 \times
  CP^3$}''},
\textsf{JHEP~0811,~089~(2008)},
\href{http://arXiv.org/abs/0807.4400}{\texttt{0807.4400}}.
%
%%CITATION = 0807.4561;%%
\bibitem{Krishnan:2008zs}
C.~Krishnan,
\textit{``{$AdS_4/CFT_3$ at One Loop}''},
\textsf{JHEP~0809,~092~(2008)},
\href{http://arXiv.org/abs/0807.4561}{\texttt{0807.4561}}.
%
%%CITATION = 0809.4038;%%
\bibitem{McLoughlin:2008he}
T.~McLoughlin, R.~Roiban and A.~A.~Tseytlin,
\textit{``{Quantum spinning strings in $AdS_4 x\times CP^3$: testing the Bethe
  Ansatz proposal}''},
\textsf{JHEP~0811,~069~(2008)},
\href{http://arXiv.org/abs/0809.4038}{\texttt{0809.4038}}.
%
%%CITATION = 0807.4897;%%
\bibitem{Gromov:2008fy}
N.~Gromov and V.~Mikhaylov,
\textit{``{Comment on the Scaling Function in AdS4 x CP3}''},
\href{http://arXiv.org/abs/0807.4897}{\texttt{0807.4897}}.
%
%%CITATION = NLIN/0003029;%%
\bibitem{Derkachov:2000ne}
S.~E.~Derkachov, D.~Karakhanian and R.~Kirschner,
\textit{``{Heisenberg spin chains based on sl(2|1) symmetry}''},
\textsf{Nucl.~Phys.~B583,~691~(2000)},
\href{http://arXiv.org/abs/nlin/0003029}{\texttt{nlin/0003029}}.
%
%%CITATION = HEP-TH/0610332;%%
\bibitem{Belitsky:2006cp}
A.~V.~Belitsky, S.~E.~Derkachov, G.~P.~Korchemsky and A.~N.~Manashov,
\textit{``{Baxter Q-operator for graded SL(2|1) spin chain}''},
\textsf{J.~Stat.~Mech.~0701,~P005~(2007)},
\href{http://arXiv.org/abs/hep-th/0610332}{\texttt{hep-th/0610332}}.
%
%%CITATION = HEP-TH/0610332;%%
\bibitem{Links}
J.~Links and A.~Foerster,
\textit{``{Integrability of a t-J model with impurities}''},
\textsf{J.~Phys.~A.~32,~147~(1999)},
\href{http://arXiv.org/abs/cond-mat/9806129}{\texttt{cond-mat/9806129}}.
%
%%CITATION = HEP-TH/0610332;%%
\bibitem{Abad}
J.~Abad and M.~R\'ios,
\textit{``Excitations and s-matrix for su(3) spin chain combining ${3}$ and
  ${3^{*}}$''},
\textsf{J.~Phys.~A.~32,~3535~(1999)},
\href{http://arXiv.org/abs/cond-mat/9806106}{\texttt{cond-mat/9806106}}.
%
%%CITATION = COND-MAT/0501197;%%
\bibitem{Essler:2005ag}
F.~H.~L.~Essler, H.~Frahm and H.~Saleur,
\textit{``{Continuum Limit of the Integrable sl(2/1) 3-$\bar{3}$ Superspin
  Chain}''},
\textsf{Nucl.~Phys.~B712,~513~(2005)},
\href{http://arXiv.org/abs/cond-mat/0501197}{\texttt{cond-mat/0501197}}.
%
%%CITATION = HEP-TH/0307015;%%
\bibitem{Beisert:2003jj}
N.~Beisert,
\textit{``{The complete one-loop dilatation operator of N = 4 super Yang-Mills
  theory}''},
\textsf{Nucl.~Phys.~B676,~3~(2004)},
\href{http://arXiv.org/abs/hep-th/0307015}{\texttt{hep-th/0307015}}.
%
%%CITATION = HEP-TH 0307042;%%
\bibitem{Beisert:2003yb}
N.~Beisert and M.~Staudacher,
\textit{``The {$\mathcal{N}=\mathord{}$4} sym integrable super spin chain''},
\textsf{Nucl.~Phys.~B670,~439~(2003)},
\href{http://arXiv.org/abs/hep-th/0307042}{\texttt{hep-th/0307042}}.
%
%%CITATION = 0901.0411;%%
\bibitem{Zwiebel:2009vb}
B.~I.~Zwiebel,
\textit{``{Two-loop Integrability of Planar N=6 Superconformal Chern- Simons
  Theory}''},
\href{http://arXiv.org/abs/0901.0411}{\texttt{0901.0411}}.
%
%%CITATION = HEP-TH 0303060;%%
\bibitem{Beisert:2003tq}
N.~Beisert, C.~Kristjansen and M.~Staudacher,
\textit{``The dilatation operator of {$\mathcal{N}=\mathord{}$4} conformal
  super yang-mills theory''},
\textsf{Nucl.~Phys.~B664,~131~(2003)},
\href{http://arXiv.org/abs/hep-th/0303060}{\texttt{hep-th/0303060}}.
%
%%CITATION = JMAPA,29,2367;%%
\bibitem{Gunaydin:1988kz}
M.~Gunaydin and S.~J.~Hyun,
\textit{``{Unitary lowest weight representations of the noncompact supergroup
  $OSp(2-n|2-m,R)$}''},
\textsf{J.~Math.~Phys.~29,~2367~(1988)}.
%
\end{thebibliography}

\end{document}